# Crystal plasticity modeling of non-Schmid yield behavior: from Ni$_3$Al single crystals to Ni-based superalloys

Devraj Ranjan[a], Sankar Narayanan[b], Kai Kadau[c], Anirban Patra[a]*

[a] Department of Metallurgical Engineering and Materials Science, Indian Institute of Technology Bombay, Mumbai, India
[b] Corporate Technology, Siemens Technology and Services Private Limited, Bangalore, India
[c] Siemens Energy Inc., Charlotte, NC, USA

* Corresponding author: anirbanpatra@iitb.ac.in

**Abstract**

A Crystal Plasticity Finite Element (CPFE) framework is proposed for modeling the non-Schmid yield behavior of L1$_2$ type Ni$_3$Al crystals and Ni-based superalloys. This framework relies on the estimation of the non-Schmid model parameters directly from the orientation- and temperature-dependent experimental yield stress data. The inelastic deformation model for Ni$_3$Al crystals is extended to the precipitate ($\gamma'$) phase of Ni-based superalloys in a homogenized dislocation density based crystal plasticity framework. The framework is used to simulate the orientation- and temperature-dependent yield of Ni$_3$Al crystals and single crystal Ni-based superalloy, CMSX-4, in the temperature range 260 – 1304 K. Model predictions of the yield stress are in general agreement with experiments. Model predictions are also made regarding the tension-compression asymmetry and the dominant slip mechanism at yield over the standard stereographic triangle at various temperatures for both these materials. These predictions provide valuable insights regarding the underlying (orientation- and temperature-dependent) slip mechanisms at yield. In this regard, the non-Schmid model may also serve as a standalone analytical model for predicting the yield stress, the tension-compression asymmetry and the underlying slip mechanism at yield as a function of orientation and temperature.

## 1. Introduction

Ni-based superalloys exhibit superior high temperature properties like high strength, high toughness, relatively high creep and thermal fatigue resistance (Geddes et al., 2010). Further, they exhibit superior resistance to degradation in oxidizing environments (Geddes et al., 2010). Given these favorable properties, Ni-based superalloys are widely used in the hot sections of gas turbine engines and nuclear power plants (Pollock and Tin, 2006). These superalloys generally consist of two phases: a Ni rich solid solution with face centered cubic (fcc) crystal structure (γ phase) and a coherent L12-type ordered intermetallic $Ni_3Al$ precipitates (γ′ phase) (Murakumo et al., 2004). While the γ phase exhibits a deformation behavior generally expected of most metals and alloys, i.e., decrease in strength with increase in temperature, the two phase alloy owes its superior thermomechanical properties at elevated temperatures to the anomalous deformation behavior of the γ′ phase, where an increase in yield strength is observed with increase in temperature up to a critical value, after which the strength decreases (Pope and Ezz, 1984; Shenoy et al., 2008).

Westbrook (1957) first observed this anomalous variation in hardness with temperature for the γ′ phase, which was later also corroborated with the findings of (Flinn, 1960) and (Davies and Stoloff, 1965), showing a peak in yield stress with increase in temperature. The γ′ phase, with $L1_2$ crystal structure, exhibits asymmetric, orientation- and temperature-dependent yield behavior in tension and compression. This type of anomalous behavior is also exhibited by various alloy forms with the $L1_2$ crystal structure, like $Ni_3Ga$ (Takeuchi and Kuramoto, 1973, 1971), $Ni_3Ge$ (Pak et al., 1977), $Ni_3Si$ (Thornton et al., 1970) and $Fe_3Al$ (Alkan and Sehitoglu, 2017a) and also B2 type NiTi crystals (Alkan and Sehitoglu, 2017b). One of the reasons contributing to this observed effect is the deviation from the widely accepted Schmid law (Schmid and Boas, 1950) for crystalline metals. Schmid law states that plastic deformation (via slip/ dislocation glide) commences only after the resolved shear stress on a slip system (Schmid stress) reaches a critical value. In the aforementioned materials exhibiting anomalous yield behavior, 'non-Schmid' yield behavior is observed. Specifically, non-Schmid stresses (along other 'non-glide' planes) also contribute to the initiation of plastic deformation, in addition to the Schmid stress (Paidar et al., 1984; Qin and Bassani, 1992). Moreover, the yield stress is generally found to increase with temperature up to about 1000 K, beyond which it decreases again (Staton-Bevan and Rawlings, 1975).

On a side note, non-Schmid yield behavior has also been studied in various bcc metals: Fe and steels (Lim et al., 2015; Spitzig and Keh, 1970), Mo (Gröger et al., 2008; Lim et al., 2013; Vitek et al., 2004; Weinberger et al., 2012), Nb and Nb-Mo (Christian, 1983; Statham et al., 1970), Ta, Ta-W and Ta-Re alloys (Cho et al., 2018; Knezevic et al., 2014) and W (Cereceda et al., 2016; Stukowski et al., 2015).

These phenomena are generally well understood in terms of the atomistic mechanisms for Ni$_3$Al (Gorbatov et al., 2016; Mryasov et al., 2002; Paidar et al., 1984; Vitek et al., 1996). Further, continuum crystal plasticity models have also been developed to address the orientation- and temperature-dependent non-Schmid yield behavior of single Ni-based superalloys (Allan, 1995; Choi et al., 2007, 2005; Cuitiño and Ortiz, 1993; Dao and Asaro, 1993; Keshavarz et al., 2016; Keshavarz and Ghosh, 2015; Leidermark et al., 2009; Österle et al., 2000; Qin and Bassani, 1992; Tinga et al., 2010; Vattré and Fedelich, 2011). Non-Schmid stresses have also been incorporated in (Ghorbanpour et al., 2017; Shenoy et al., 2008) to model the deformation of polycrystalline Ni-based superalloys. While these crystal plasticity studies have incorporated non-Schmid stresses, their effect on the polycrystalline response has not been explicitly isolated from other factors (back stress) that may have contributed to the tension-compression asymmetry and the orientation dependent response. Moreover, non-Schmid yield behavior is not the main focus of these studies. Generally speaking, the non-Schmid model parameters in the above mentioned works are either adapted from lower scale simulations or estimated via (trial and error) fit to the experimental stress-strain response. While these are viable strategies, an alternative and perhaps more appealing strategy might be where the model parameters pertaining to the non-Schmid yield behavior are analytically estimated from the experimental yield stress data, without the need for running calibration simulations. This is especially important given the highly anisotropic orientation-dependent response of the material under consideration, as well as the variation in response with temperature. In this work, we develop a framework and demonstrate a method to address these issues.

This work presents a continuum crystal plasticity framework for modeling the non-Schmid yield behavior of Ni$_3$Al crystals and Ni-based superalloys. We first describe a method for estimating the non-Schmid model parameters directly from available experimental data. This method relies on the non-Schmid yield formulation originally proposed by (Qin and Bassani, 1992). Here, we extend the framework to model Ni$_3$Al crystals, with a detailed algorithm for estimating the non-Schmid model parameters for Ni$_3$Al from the experimental orientation- and temperature-dependent yield stress data. In the latter part of this manuscript, we describe the implementation of the non-Schmid model in a homogenized Crystal Plasticity Finite Element (CPFE) framework that uses iso-stress assumption (as a first order approximation) to model deformation in the two phases. The framework is then used to predict the yield behavior of two materials: Ni$_3$Al crystals and Ni-based superalloy, CMSX-4. We present the CPFE model predictions of the yield stress, tension-compression asymmetry, and the dominant slip mechanism at yield, for a range of loading orientations and temperatures and compare with relevant experiments.

This framework may be extended to study the non-Schmid yield behavior of other L1$_2$ crystals and Ni-based superalloys, with minor modifications.

**2. Current understanding of the non-Schmid yield behavior of Ni$_3$Al**

In Ni$_3$Al, the $<\bar{1}10>$ super-dislocation gets dissociated into two super partials separated by an anti-phase boundary (APB), as shown in figure 1(a) (Flinn, 1960; Yamaguchi et al., 1982). The APB energy on the $\{111\}$ plane is relatively high to allow slip of a single $1/2<\bar{1}10>$ partial dislocation. Further, movement of the $\{111\}$ layer by another $1/2<\bar{1}10>$ restores order. Consequently, the $1/2<\bar{1}10>$ partials must travel in pairs separated by APBs (Koehler and Seitz, 1947). These two $1/2<\bar{1}10>$ super partials may in turn dissociate into $1/6<11\bar{2}>$ Shockley partials separated by Complex Stacking Faults (CSFs) similar to an fcc crystal (Kruml et al., 2002). A dislocation structure in L1$_2$ crystals therefore consists of two pairs of Shockley partials each separated by a CSF, and separated from each other by an APB (Copley and Kear, 1967).

The APB energy associated with the $\{001\}$ plane is lower than the $\{111\}$ plane, and this acts as the driving force for the cross-slip of super partials from the $\{111\}$ primary octahedral slip plane onto a $\{001\}$ cube plane (Collins and Stone, 2014). However, an extended dislocation core is constrained to glide in the $\{111\}$ plane of its fault (Crimp, 1989). Although an extended screw dislocation cannot cross-slip, it may form a constriction and then be free to move to other planes ($\{001\}$ or another $\{111\}$ plane) (Thornton et al., 1970). Such dislocations may further cross-slip onto a secondary $\{111\}$ plane. The APB however resides on the $\{001\}$ plane, thus leaving the dislocation core extended onto three planes: primary octahedral, cube and secondary octahedral planes, and this causes pinning of the dislocation segments. Extension of the dislocation core onto multiple planes is of course orientation-dependent and the dislocation core spreading onto a cube plane is not geometrically feasible for certain orientations. The major reason for the hardening and anomalous yield behavior in these crystals is the hardening caused by these pinned segments of screw dislocations for which driving force is the difference in APB energy on the $\{111\}$ and $\{001\}$ planes (Paidar et al., 1984). This pinning of partials is generally referred as Kear Wilsdorf (KW) lock formation in the literature (Copley and Kear, 1967; Wang-Koh, 2017). These mechanisms are schematically shown in figure 1(a). An increase in temperature promotes cross-slip, thus increasing the density of pinned segments and the hardening caused by these segments. On the other hand, the strength of the pinned segments decreases with temperature. The combination of these two mechanisms contributes to the so called anomalous behavior. To add to the complexity of the yield phenomena, a secondary mechanism that is responsible for anomalous behavior is the activation of cube slip systems; cube planes are not close packed planes in these crystals (Österle et al., 2000). Activation of cube slip along the {001} planes is observed at $T > 1000$ K for certain crystal orientations, in addition to octahedral slip (Thornton et al., 1970). Above this critical temperature, dislocations on cube planes are found to be of both screw and edge dislocation types without any cross-slip (Lall et al., 1979; Paidar et al., 1984).

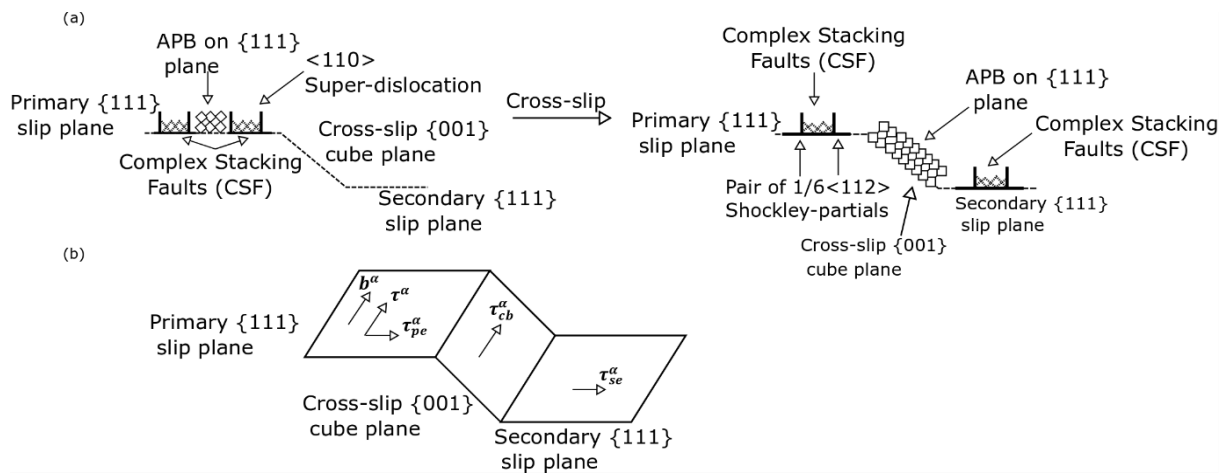

**Figure 1.** Schematic of (a) formation of partials, cross-slip and Kear Wilsdorf lock formation, and (b) Schmid and non-Schmid slip systems in Ni$_3$Al.

These crystallographic defects have been studied using a combination of atomistic simulations (Gorbatov et al., 2016; Mryasov et al., 2002; Paidar et al., 1984; Vitek et al., 1996) and experimental microscopy (Lall et al., 1979). Ab-initio simulations have been performed to calculate the Anti-Phase Boundary Energy (APBE) associated with the defect structures (Gorbatov et al., 2016; Mryasov et al., 2002; Yamaguchi et al., 1982). A model correlating the APBEs and the resolved shear stresses on the non-Schmid planes with the yield phenomenon was proposed by Paidar et al. (1984). Further, model predictions were also compared with the dislocation characteristics observed from experiments (Lall et al., 1979). We briefly summarize their findings in the following. The stresses which cause the Shockley partials to constrict are the shear stresses resolved on the primary ($p$) and secondary ($s$) {111} slip planes in the edge ($e$) direction, $\tau_{pe}$ and $\tau_{se}$. These slip planes are shown schematically in figure 1(b). When applied along one direction (say tension or compression), these resolved stresses constrict the Shockley partials, promoting cross-slip. When the direction of the applied stress is reversed, $\tau_{pe}$ and $\tau_{se}$ extend the Shockley partials, thus hindering cross-slip. Therefore, in one direction, the effect of the stresses is to increase the yield stress, and, in the opposite direction, the effect is to decrease it, causing the observed tension-compression asymmetry (Copley and Kear, 1967). Note that while the APB is on the {001} plane (plane with the lowest energy), the dissociation into Shockley partials can occur only on the {111} planes, thus making the core of the dislocation non-planar (Allan, 1995; Keshavarz and Ghosh, 2015). In addition to $\tau_{pe}$ and $\tau_{se}$, cross slip is also promoted by a third non-Schmid shear stress, $\tau_{cb}$, the stress resolved on the cube ($c$) cross-slip plane in the direction of the Burgers ($b$) vector, $\tau_{cb}$ (Paidar et al., 1984).

This mechanism was proposed as a constitutive model by (Paidar et al., 1984), where the activation energy of cross-slip from {111} to {001} was given as a function of APB energies, resolved shear stresses on {111} and {001} planes, and the width of dislocation cores on primary and secondary slip planes. Accordingly, the enthalpy for cross-slip proposed by (Cuitiño and Ortiz, 1993; Paidar et al., 1984) is of the form:

$$H_c = \frac{\mu b^3}{4\pi} \left\{ h + c + \kappa_1(\tau_{pe} - \kappa_2 \tau_{se}) - \sqrt{\left(\frac{1}{\sqrt{3}} - \frac{\Gamma^{\{001\}}}{\Gamma^{\{111\}}} + |\tau_{cb}|\right) \frac{b}{B}} \right\} \qquad (1)$$

where $h, c, \kappa_1,$ and $\kappa_2$ are material constants, $\mu$ is the Shear modulus, $b$ is the Burgers vector magnitude, while $\Gamma^{\{001\}}$ and $\Gamma^{\{111\}}$ are the APBEs per unit area on the cube and octahedral planes, considered to be independent of temperature and $B = \mu b^2/2\pi \, \Gamma^{\{111\}}$ represents the equilibrium (stress-free) spacing of the super partials on the {111} plane. The critical resolved shear stress (CRSS) for plastic deformation was given as function of this enthalpy, phenomenologically representative of the density of the cross-slip dislocations (Allan, 1995), i.e.,

$$\tau_{cr}^\alpha = A \exp\left(-\frac{H_c}{3kT}\right) \qquad (2)$$

where $A$ is representative of the hardening coefficient, $k$ is the Boltzmann's constant, and $T$ is the absolute temperature. This model explains the fact that the CRSS is both orientation- and temperature-dependent. This model was implemented in a crystal plasticity constitutive framework by (Allan, 1995) and later by (Keshavarz and Ghosh, 2015) to model the tension-compression asymmetry of Ni$_3$Al single crystals. Moreover, cube slip was also considered using a temperature-dependent CRSS function (calibrated to experimental data) to predict the yield behavior over various orientations of the standard stereographic triangle. Although the models proposed by (Allan, 1995; Keshavarz and Ghosh, 2015) were able to capture the observed tension-compression asymmetry fairly well, calibration of the non-Schmid model parameters itself would presumably require some trial and error. Specifically, the nature of the non-linear constitutive equations used in these works are such that direct estimation of the non-Schmid model parameters for yield from the experimental data is not entirely intuitive. In order to directly estimate model parameters from the experimental data, a linear non-Schmid model form may be preferable. Note that a variant of the CRSS model in eq. (2) was also proposed by (Cuitiño and Ortiz, 1993) to model the orientation- and temperature-dependent yield.

Qin and Bassani (1992) proposed a non-Schmid yield criterion for L1$_2$ crystals as a linear combination of the resolved shear (Schmid) stress and non-Schmid stresses as follows:

$$\tau_0 = \tau^\alpha + \sum_{i=1}^{N_{ns}} a_i \tau_i^\alpha \tag{3}$$

where, $\tau^\alpha$ is the resolved shear stress on the primary slip system $\alpha$, $\tau_i^\alpha$ is the $i^{th}$ non-Schmid stress component, and $a_i$ is the corresponding non-Schmid coefficient. The non-Schmid stresses are summed over all possible systems, $N_{ns}$. Yield occurs when the sum of stresses on the RHS reaches the critical value, $\tau_0$, a material parameter. Physically, the dislocation core, which is spread on different planes, needs to constrict on to the primary plane for slip to occur. In a separate work, Dao and Asaro (1993) argued that the non-Schmid formulation proposed by (Paidar et al., 1984) reduces to a similar linear form, as in eq. (3), by assuming a Taylor series expansion of eq. (2) and ignoring the higher order stress terms (the contribution of the higher order stress terms are expected to be relatively small compared to 1.0). Based on this assumption, eq. (2) reduces to the following form (Dao and Asaro (1993)):

$$\tau_{cr}^\alpha = C_0 + C_1 \tau_{pe} + C_2 \tau_{se} + C_3 |\tau_{pb}| \tag{4}$$

where, $C_i = C_i(T)$ are temperature-dependent functions, for $i = 0,1,2,3$.

We have used this linear non-Schmid yield formulation for modeling the yield behavior of Ni$_3$Al crystals in the present work.

## 3. Estimation of non-Schmid coefficients for Ni₃Al from experimental data

We have systematically analyzed the experimental yield stress data for Ni$_3$Al single crystals (Heredia and Pope, 1991) for different orientations and temperatures. The yield stress in these crystals is clearly anisotropic and the Schmid factor alone cannot account for the observed anisotropy. Further, it is evident that the yield stress depends on the type of loading (tension or compression), which activate a different set of slip systems.

{111} is the primary slip plane for octahedral slip in Ni$_3$Al crystals. Both forward and backward slip is considered by assuming positive and negative slip directions as different slip systems in order to account for the type/sense of loading on the yield stress. Thus, we have 24 octahedral {111}<110> slip systems now, in place of the 12 possible combinations of the octahedral slip systems. These slip systems are numbered 1 through 24 in table A1 in the appendix. Further, the additional 6 cubes {100}<110> slip systems makes the total number of slips systems as 30. These slip systems are also listed in table A1. The Qin-Bassani yield formulation (eq. (3)) may be rewritten, in terms of the resolved shear stresses from the Schmid and non-Schmid slip systems, as

$$\tau_{0,oct}^\alpha = \boldsymbol{m}^\alpha \cdot \boldsymbol{\sigma} \cdot \boldsymbol{n}^\alpha + a_1 \boldsymbol{m}_{pe}^\alpha \cdot \boldsymbol{\sigma} \cdot \boldsymbol{n}_{pe}^\alpha + a_2 \boldsymbol{m}_{se}^\alpha \cdot \boldsymbol{\sigma} \cdot \boldsymbol{n}_{se}^\alpha + a_3 \boldsymbol{m}_{cb}^\alpha \cdot \boldsymbol{\sigma} \cdot \boldsymbol{n}_{cb}^\alpha$$

$$= \tau^\alpha + a_1 \tau_{pe}^\alpha + a_2 \tau_{se}^\alpha + a_3 \tau_{cb}^\alpha \qquad (5)$$

where, $\boldsymbol{\sigma}$ is the applied stress tensor, $(\boldsymbol{m}^\alpha, \boldsymbol{n}^\alpha)$ are the unit vector corresponding to the slip and slip plane normal directions on the primary slip system $\alpha$, $(\boldsymbol{m}_{pe}^\alpha, \boldsymbol{n}_{pe}^\alpha)$, $(\boldsymbol{m}_{se}^\alpha, \boldsymbol{n}_{se}^\alpha)$ and $(\boldsymbol{m}_{cb}^\alpha, \boldsymbol{n}_{cb}^\alpha)$ are the slip and slip plane normal directions associated with the primary edge, secondary edge and cube planes mentioned earlier, and $a_1$, $a_2$ and $a_3$ are the non-Schmid coefficients. Note that eq. (5) is the material- and slip system-specific form of the generalized non-Schmid formulation in eq. (3), where $\tau^\alpha$ is the Schmid stress, $\tau_{pe}^\alpha$, $\tau_{se}^\alpha$ and $\tau_{cb}^\alpha$ are the dominant non-Schmid stresses in this material. For yield to occur, the combination of stresses should reach a critical value, $\tau_{0,oct}^\alpha$. The parameter, $\tau_{0,oct}^\alpha$, is a temperature-dependent material property and independent of the loading orientation. Physically, $\tau_{0,oct}^\alpha$ is representative of the hardening caused by the cross-slip dislocations and KW locks mentioned in the previous section. A parametric study of the effect of the non-Schmid coefficients, $a_1$, $a_2$ and $a_3$, on the tension-compression asymmetry resulting from octahedral slip is discussed in Appendix B.

If the experimental yield stress is known for a certain loading orientation, the corresponding Schmid and non-Schmid resolved shear stresses in eq. (5) can be calculated analytically. If there are experimental data for multiple loading orientations at any temperature, a system of linear equations may be written, with each equation corresponding to a specific loading orientation. The solution of these system of linear equations may provide the values of the unknowns: $a_1$, $a_2$, $a_3$ and $\tau_{0,oct}^{\alpha}$ for a specific temperature.

We have used data from a single set of experiments reported in (Heredia and Pope, 1991). An example data set from the experiments at 283 K is shown in Table 1. The experimental yield stress data has not been reported at the exact same temperatures for the different loading orientations. As a first order approximation, the nominal values for the different temperatures at which these experiments have been reported are binned to a mean temperature value in the range $\pm 20$ K (the implicit assumption being that the non-Schmid model parameters do not change in this temperature range). Only four set of linear equations are required to calculate these four unknowns. However, for most temperatures, there are more experimental data points due to a combination of different loading orientations and the type of loading (tension and compression). This may lead to an over constrained system of equations with *non-unique* solutions.

Initially, we tried solving these equations using a least square regression fit method for each temperature. Model parameters obtained from this method are shown in figures A2 and A3 in the appendix. While this did provide us with the temperature-dependent values of $a_1$, $a_2$, $a_3$ and $\tau_{0,oct}^{\alpha}$, clearly no trends (monotonic or otherwise) were observed as a function of temperature. This could be due to the earlier mentioned fact that this is an over constrained system of equations. Moreover, even small errors in measurements or digitization of the experimental data may lead to non-physical solutions of the non-Schmid parameters, for example, very large values of these material parameters. Physically, the parameters $a_1$, $a_2$, and $a_3$ are expected to be temperature dependent functions obtained by Taylor series expansion of the coefficients (cf. eq. (4)) (Dao and Asaro, 1993). Hence, it is perhaps reasonable to expect some trends with respect to temperature. Finally, we also note that neglect of higher order stress terms in the linear form of the non-Schmid model may also contribute to this discrepancy.

In order to get around this issue, we have employed constrained fitting of the non-Schmid parameters. These are motivated from the considerations made earlier in (Patra et al., 2014), where the experimental data was used to estimate the temperature-dependent non-Schmid coefficients of bcc-Fe. In this earlier work, the non-Schmid coefficients were estimated in two steps: one of the non-Schmid coefficients, $a_1$, was first calculated using least squares regression, and then the other coefficients were estimated by solving a system of linear equations at each temperature. While this method worked reasonably well to predict the non-Schmid parameters in the earlier work, here we have tried to (a) estimate all the non-Schmid model parameters simultaneously, and (b) develop a more general algorithm that may also be applied to other materials for estimating non-Schmid model parameters. The constraints and the algorithm for determining the non-Schmid coefficients are described in the following:

(a) The non-Schmid coefficients, $a_1$, $a_2$, $a_3$, are constrained to values in the range $[-1,1]$. The assumption behind this constraint is that the contribution of the non-Schmid stress to the CRSS at any given temperature should not be more than that of the resolved shear stress on the primary slip plane, i.e., the Schmid stress. Absolute value of the non-Schmid coefficients generally used in prior works are also less than 1 (Dao and Asaro, 1993; Keshavarz and Ghosh, 2015). We generate all possible combinations of $a_1$, $a_2$, and $a_3$ in the range $[-1,1]$ with an interval size of $0.02$.

(b) For a given temperature, the experimental yield stress data for each loading orientation and type of loading is used to calculate the corresponding Schmid and non-Schmid stress components (eq. 5). These stress components are decided based on the Schmid factor of the "active" slip system out of the 30 available slip systems. Physically, the onset of slip can be associated with the slip system having the highest Schmid factor. The same assumption is followed here, *albeit* with appropriate consideration of the sign of the shear stress components in tension and compression (also see appendix B).

(c) The non-Schmid coefficients are multiplied with the respective non-Schmid stresses and added to the Schmid stress to obtain a guess value of $\tau_{0,oct}^{\alpha}$. Note that the stress tensor, $\boldsymbol{\sigma}$, in eq. (5) has only one component along the direction of loading (direction 3 in our case) for uniaxial deformation.

(d) For a set of non-Schmid coefficients, $a_1$, $a_2$, and $a_3$, the mean predicted value of $\tau_0^{\alpha}$ is calculated and the corresponding normalized root mean square deviation ($RMSD$) of the predicted $\tau_0^{\alpha}$ (across all orientations and types) from the mean value is calculated as follows

$$RMSD = \sqrt{\frac{1}{N}\sum_{i=1}^{N}\left(\frac{\tau_0^\alpha|_i - \langle\tau_0^\alpha\rangle}{\langle\tau_0^\alpha\rangle}\right)^2} \quad (6)$$

where $\tau_0^\alpha|_i$ is the predicted CRSS value for the $i^{th}$ loading orientation/type, $\langle\tau_0^\alpha\rangle$ is the mean predicted CRSS, and $N$ is the total number of loading orientations/type. Ideally, $RMSD$ should be equal to zero. Our goal is to find values of $a_1$, $a_2$, $a_3$, and $\tau_{0,oct}^\alpha$ that minimize the $RMSD$.

(e) For a set of non-Schmid coefficients, $a_1$, $a_2$, $a_3$, the maximum allowable $RMSD$ should not be more than 0.12 (or 12%). Based on the analysis of the experimental data, we have observed that $RMSD$ beyond this value generally leads to erroneous solutions. At this point, multiple possible feasible sets of $a_1$, $a_2$, and $a_3$ are obtained that satisfy the $RMSD$ criteria across all loading orientations/type.

(f) The most appropriate set of $a_1$, $a_2$, and $a_3$ at any given temperature is then chosen subject to the condition that the $RMSD$ is minimum and the parameters, $a_1$, $a_2$, and $a_3$, follow a monotonic trend with respect to temperature. This monotonic trend is enforced at all temperatures, except for the temperature range in which a reversal of the yield stress occurs ($\sim 800 - 1000$ K).

This constitutes the general algorithm for determination of the non-Schmid model parameters as a function of temperature. Except for the final step, no manual intervention is needed during the execution of this algorithm. While this algorithm does not provide a unique set of parameters, it may provide one which is physically justifiable. This algorithm has been executed for all the available experimental data given in (Heredia and Pope, 1991). We note that a regression based approach was also used in (Österle et al., 2000) for determining the non-Schmid coefficients of Ni-based superalloys. Finally, we also note that this data-driven approach for parameter estimation does not have much connection to the underlying physics. More physically based approaches that rely on minimization of the energy required for the dislocation to glide have also been used elsewhere (Alkan and Sehitoglu, 2017b) to calculate the CRSS for non-Schmid yield, with knowledge of the core structure of the dislocations from atomistic simulations.

An example application of this algorithm to calculate the non-Schmid parameters for Ni$_3$Al at 283 K is given in Table 1. In this table, the Schmid and non-Schmid stresses are calculated directly from the experimental yield stress. This table also shows the predicted CRSS value for the $i^{th}$ loading orientation/type, $\tau_0^\alpha|_i$, and the mean predicted CRSS $\langle\tau_0^\alpha\rangle$ for the non-Schmid coefficients, $a_1$, $a_2$, and $a_3$ with the lowest $RMSD$. A similar procedure is followed at all temperatures to obtain the non-Schmid parameters from the experimental data.

**Table 1.** Estimation of non-Schmid coefficients of Ni$_3$Al from the experimental yield stress data (Heredia and Pope, 1991) at 283 K. T indicates tension and C indicates compression.

| Loading Orientation/ Type | Exp. YS (MPa) | $\tau^\alpha$ (MPa) | $\tau^\alpha_{pe}$ (MPa) | $\tau^\alpha_{se}$ (MPa) | $\tau^\alpha_{cb}$ (MPa) | $a_1$, $a_2$, $a_3$ | $\tau^\alpha_0\vert_i$ (MPa) | $\langle\tau^\alpha_0\rangle$ (MPa) | $RMSD$ |
|---|---|---|---|---|---|---|---|---|---|
| [0 0 1]/ T | 168 | 68.586 | 39.598 | 39.598 | 0 | -0.04, -0.02, -0.02 | 66.209 | 68.123 | 0.068 |
| [0 0 1]/ C | 168 | 68.586 | -39.598 | -39.598 | 0 | | 70.961 | | |
| [$\bar{1}$ 9.1 10.1]/ T | 162 | 71.902 | -34.033 | 0 | 62.269 | | 72.018 | | |
| [$\bar{1}$ 9.1 10.1]/ C | 164 | 72.789 | 34.453 | 0 | 63.037 | | 70.151 | | |
| [$\bar{1}$ 1.1 1.2]/ T | 210 | 67.176 | -35.258 | -29.291 | 98.452 | | 67.203 | | |
| [$\bar{1}$ 1.1 1.2]/ C | 190 | 60.778 | 31.901 | 26.502 | 89.076 | | 57.191 | | |
| [$\bar{1}$ 3.2 16]/ T | 159 | 75.151 | 21.949 | 35.412 | 22.886 | | 72.648 | | |
| [$\bar{1}$ 3.2 16]/ C | 143 | 67.589 | -19.741 | -31.849 | 20.583 | | 68.604 | | |

Finally, the non-Schmid parameters and CRSS are expressed as piecewise polynomial functions of temperature. These polynomial functions are given in Table 2. These equations are valid in the temperature range of $260 - 1304$ K, since the experimental data was available only in this range. The obtained data points and the corresponding polynomial functions are plotted for $a_1$, $a_2$, and $a_3$ in figure 2, and for $\tau^\alpha_{0,oct}$ in figure 3(a).

The trends for $a_1, a_2, a_3$ and $\tau^\alpha_{0,oct}$ are noticeably different in two temperature regimes: $T > 1000$ K and $T < 1000$ K. Generally speaking, the values of $a_1$, $a_2$, and $a_3$ are monotonically decreasing, and the value of $\tau^\alpha_0$ is monotonically increasing in the regime: $T < 1000$ K. There is an inversion in the observed trend above 1000 K. These trends hint towards the dominance of KW lock formation governing the deformation mechanism below 1000 K. As described earlier in section 2, KW lock formation depends on the propensity of cross-slip, which in turn is a thermally activated process. As the temperature increases, the number of cross-slip dislocations and KW locks thus increase. Beyond a certain temperature (~1000 K), the strength (resistance to dislocation glide) of these locks starts decreasing and hence the observed decrease in $\tau^\alpha_0$. It is perhaps difficult to *directly* validate these non-Schmid parameters with respect to the underlying physics. However, as will be seen later (section 5), model predictions of the yield stress and underlying slip system activity (where available) compare favorably with the experimental observations. This may serve as an indirect validation for the estimated model parameters.

It should be noted that both these competing mechanisms were modeled earlier by (Allan, 1995; Keshavarz and Ghosh, 2015). They modeled the cross-slip dislocations using eqs. (1) and (2), while the strength coefficient of these dislocations/locks was modeled using a phenomenological term (that decreases with increasing temperature) fit to the experimental data.

In our work, once the parameters, $a_1, a_2, a_3$ and $\tau_{0,oct}^{\alpha}$, are estimated directly from the experimental data, they may be directly used in a crystal plasticity finite element model (described in the following sections) or even an analytical model to predict the yield stress as a function of orientation and temperature. This is an advantage of using the linear form of the non-Schmid yield formulation over the non-linear form used in the above mentioned works; a non-linear model may not be amenable to analytical calculations as easily.

**Table 2.** Non-Schmid parameters and CRSS as a function of absolute temperature for Ni$_3$Al.

| Parameter | Value | Range |
|---|---|---|
| $a_1$ (eq. (7)) | $-4.411 \times 10^{-12}T^5 + 7.927 \times 10^{-9}T^4 - 5.609 \times 10^{-6}T^3 + 1.952 \times 10^{-3}T^2 - 3.344 \times 10^{-1}T + 2.256 \times 10^1$ | $260 \text{ K} \leq T < 497 \text{ K}$ |
| | $-0.140$ | $497 \text{ K} \leq T < 706 \text{ K}$ |
| | $-4.101 \times 10^{-4}T + 1.498 \times 10^{-1}$ | $706 \text{ K} \leq T < 755 \text{ K}$ |
| | $-0.160$ | $755 \text{ K} \leq T < 849 \text{ K}$ |
| | $-2.122 \times 10^{-2}T + 1.787 \times 10^1$ | $849 \text{ K} \leq T < 854 \text{ K}$ |
| | $-0.260$ | $854 \text{ K} \leq T < 1008 \text{ K}$ |
| | $1.2757 \times 10^{-16}T^7 - 1.035 \times 10^{-12}T^6 + 3.607 \times 10^{-9}T^5 - 6.994 \times 10^{-6}T^4 + 8.150 \times 10^{-3}T^3 - 5.707\,T^2 + 2.223 \times 10^3\,T - 3.717 \times 10^5$ | $1008 \text{ K} \leq T < 1304 \text{ K}$ |
| $a_2$ (eq. (8)) | $-5.037 \times 10^{-7}T^2 + 2.127 \times 10^{-4}T - 3.987 \times 10^{-2}$ | $260 \text{ K} \leq T < 502 \text{ K}$ |
| | $-6.002 \times 10^{-2}$ | $502 \text{ K} \leq T < 572 \text{ K}$ |
| | $-1.053 \times 10^{-3}T + 5.263 \times 10^{-1}$ | $572 \text{ K} \leq T < 633 \text{ K}$ |
| | $-1.402 \times 10^{-1}$ | $633 \text{ K} \leq T < 682 \text{ K}$ |
| | $-2.632 \times 10^{-4}T + 3.947 \times 10^{-2}$ | $682 \text{ K} \leq T < 758 \text{ K}$ |
| | $-0.160$ | $758 \text{ K} < T < 1136 \text{ K}$ |
| | $-1.251 \times 10^{-3}T + 1.261$ | $1136 \text{ K} \leq T < 1168 \text{ K}$ |
| | $4.611 \times 10^{-6}T^2 - 1.287 \times 10^{-2}T + 8.545$ | $1168 \text{ K} \leq T \leq 1304 \text{ K}$ |
| $a_3$ (eq. (9)) | $-1.891 \times 10^{-13}T^5 + 3.28 \times 10^{-10}T^4$ $- 2.137 \times 10^{-7}T^3$ $+ 6.215 \times 10^{-5} T^2$ $- 7.518 \times 10^{-3} T + 0.214$ | $260 \text{ K} \leq T < 546 \text{ K}$ |
| | $-0.180$ | $546 \text{ K} \leq T < 706 \text{ K}$ |
| | $-1.640 \times 10^{-3}T + 9.790 \times 10^{-1}$ | $706 \text{ K} \leq T \leq 755 \text{ K}$ |
| | $-0.260$ | $755 \text{ K} \leq T < 849 \text{ K}$ |
| | $4.243 \times 10^{-2}T - 3.632 \times 10^1$ | $849 \text{ K} \leq T < 854 \text{ K}$ |
| | $-0.060$ | $854 \text{ K} \leq T < 1136 \text{ K}$ |
| | $7.66223256593 \times 10^{-9}T^4$ $- 3.54756643114438 \times 10^{-5}T^3$ $+ 6.15741472834139 \times 10^{-2} T^2$ $- 4.7483288242742 \times 10^1 T$ $+ 1.37266826682315 \times 10^4$ | $1136 \text{ K} \leq T < 1207 \text{ K}$ |
| | $-0.020$ | $1207 \text{ K} \leq T < 1304 \text{ K}$ |
| $\tau_{0,oct}^\alpha$ (eq. (10)) | $2.589 \times 10^{-15}T^6 - 9.269 \times 10^{-12}T^5$ $+ 1.033 \times 10^{-8}T^4$ $- 2.784 \times 10^{-6}T^3$ $- 1.452 \times 10^{-3} T^2$ $+ 9.265 \times 10^{-1} T$ $- 6.522 \times 10^1 \text{MPa}$ | $260 \text{ K} \leq T \leq 1304 \text{ K}$ |

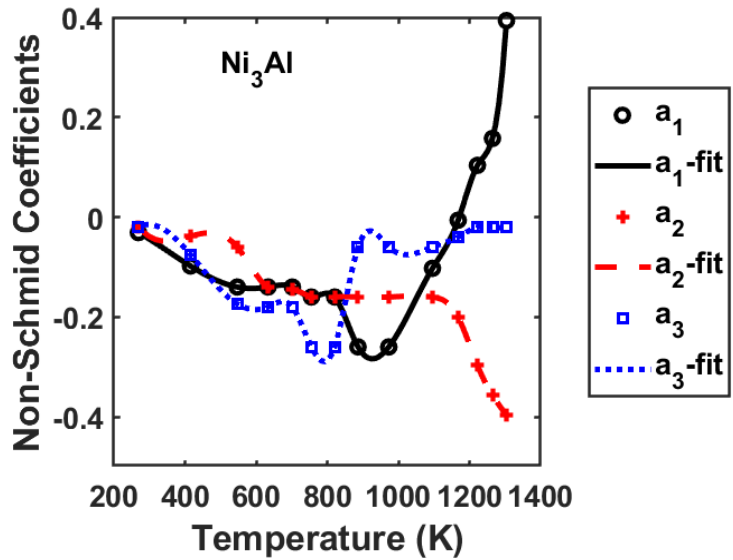

**Figure 2.** Variation of non-Schmid coefficients, $a_1$, $a_2$ and $a_3$ with temperature for octahedral slip in $Ni_3Al$.

In addition to the octahedral slip systems, there exist 6 cube slip systems, which may get activated near the $\langle 0\ 1\ 1 \rangle$ and $\langle 1\ 1\ 1 \rangle$ poles of the stereographic triangle at elevated temperatures. The temperature dependent CRSS for cube slip was earlier estimated by (Allan, 1995). We have directly used those values here. The CRSS for cube slip is given in the equation below and is also plotted in figure 3(b).

$$\tau^\alpha_{0,cube} = \begin{cases} 858 - 0.577T \text{ MPa}; & T > 915 \text{ K} \\ 330 \text{ MPa}; & T < 915 \text{ K} \end{cases} \tag{11}$$

As discussed in this and the preceding sections, there is a huge body of literature discussing the non-Schmid phenomena for octahedral slip in the L1$_2$ crystals. However, to the best of our knowledge, the same has not been reported for cube slip. Hence, we assume that non-Schmid effects are not dominant for cube slip. Further, no tension-compression asymmetry is observed at the higher temperatures for both Ni$_3$Al and CMSX-4 for the loading orientations where cube slip may be activated (based on the highest Schmid factor). For example, see the experimental data shown for the $[\bar{1}\ 9.1\ 10.1]$ and $[\bar{1}\ 1.1\ 1.2]$ loading orientations in figure 6 (for Ni$_3$Al) and $[\bar{1}\ 1\ 1]$ loading orientation in figure 9 (for CMSX-4) (cf. Section 5). Since there is no experimental proof to suggest that cube slip may be influenced by non-Schmid components and/or demonstrates tension-compression asymmetry, we have not considered non-Schmid components for cube slip.

In the following sections, we describe the implementation of these non-Schmid equations in a CPFE framework to simulate the yield behavior of both Ni$_3$Al crystals and two-phase Ni-based superalloys.

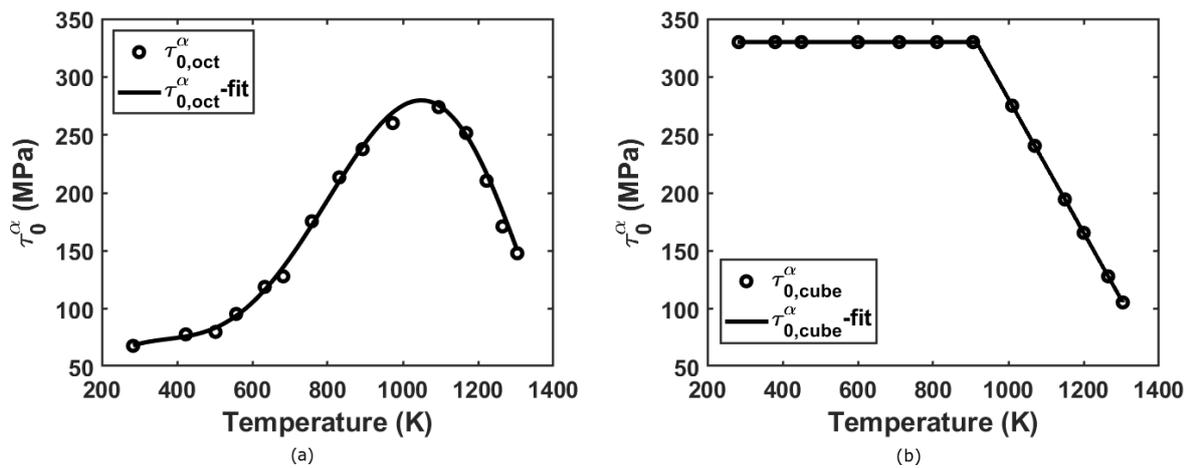

**Figure 3.** Variation of critical resolved shear stress with temperature for (a) octahedral slip, and (b) cube slip. The model for cube slip is adapted from (Allan, 1995).

## 4. Crystal plasticity finite element framework

We have implemented the non-Schmid model in a crystal plasticity constitutive framework that accounts for the homogenized deformation of two phase Ni-based superalloys under a Sachs-type iso-stress assumption (cf. (Tinga et al., 2008)).

The microstructure in the homogenized crystal plasticity framework is represented primarily in terms of the volume fraction of the $\gamma'$ phase, $f_{\gamma'}$, and the mean precipitate size, $d_{\gamma'}^{eff}$. Further, the effective precipitate size is calculated as the weighted mean of the primary ($d_{\gamma'}^p$) and secondary ($d_{\gamma'}^s$) precipitate sizes, with volume fractions, $f_{\gamma'}^p$ and $f_{\gamma'}^s$, respectively, i.e.,

$$\frac{1}{d_{\gamma'}^{eff}} = \frac{1}{f_{\gamma'}}\left(\frac{f_{\gamma'}^p}{d_{\gamma'}^p} + \frac{f_{\gamma'}^s}{d_{\gamma'}^s}\right); \; f_{\gamma'} = f_{\gamma'}^p + f_{\gamma'}^s \tag{12}$$

Depending on whether the superalloy has a unimodal or a bimodal distribution of precipitates, the appropriate terms may be neglected in eq. (12). Note that the constitutive framework may be used for modeling the deformation behavior of Ni$_3$Al crystals, with no $\gamma$ phase, simply by using $f_{\gamma'} = 1$.

### 4.1. Finite deformation kinematics

This framework is based on the multiplicative dissociation of the deformation gradient $\boldsymbol{F}$ into the elastic and inelastic parts, i.e., $\boldsymbol{F} = \boldsymbol{F}^e \cdot \boldsymbol{F}^i$ (Asaro and Rice, 1977). Here, $\boldsymbol{F}^i$ accounts for the effects of inelastic deformation on a strain free lattice, mapping the reference configuration to an intermediate configuration in which the lattice has the same orientation, and $\boldsymbol{F}^e$ accounts for the lattice distortion and the rigid body rotation that carries the intermediate isoclinic configuration to the current, deformed configuration.

The elastic Green strain tensor, $\boldsymbol{E}^e = 1/2(\boldsymbol{F}^{eT} \cdot \boldsymbol{F}^e - \boldsymbol{I})$ is related to the second Piola-Kirchoff stress in the intermediate configuration by, $\boldsymbol{S} = \boldsymbol{C}_{eff}:\boldsymbol{E}^e$, where $\boldsymbol{C}_{eff}$ is the effective fourth rank elasticity tensor of the lattice. The second order Piola-Kirchoff stress is related to the Cauchy stress, $\boldsymbol{\sigma}$, in the current configuration: $\boldsymbol{S} = det(\boldsymbol{F}^e)\boldsymbol{F}^{e^{-1}} \cdot \boldsymbol{\sigma} \cdot \boldsymbol{F}^{e^{-T}}$. For a homogenized material, the individual components of the effective elasticity tensor are given by

$$C_{eff}^{ijkl} = \left((1 - f_{\gamma'})C_m^{ijkl-1} + f_{\gamma'}C_p^{ijkl-1}\right)^{-1} \tag{13}$$

where, $C_m^{ijkl}$ is the elasticity tensor component of the matrix ($\gamma$) phase, and $C_p^{ijkl}$ is the elasticity tensor component of the precipitate ($\gamma'$) phase, and the superscripts, $i, j, k, l$ refer to the respective indices. Throughout this manuscript, the subscripts $m$ and $p$ in lowercase letters are used to denote the *matrix* and the *precipitate* phases, respectively.

The evolution of the inelastic deformation gradient is given by: $\dot{\boldsymbol{F}}^i = \boldsymbol{L}^i \cdot \boldsymbol{F}^i$, where, $\boldsymbol{L}^i$ is the inelastic velocity gradient. Under an iso-stress assumption, $\boldsymbol{L}^i$ is the tensor sum of the crystallographic shearing rates over all possible slip systems in the material, i.e.,

$$\boldsymbol{L}^i = (1 - f_{\gamma'}) \sum_{\alpha=1}^{N_m} \dot{\gamma}_m^\alpha \, \boldsymbol{m}_{0m}^\alpha \otimes \boldsymbol{n}_{0m}^\alpha + f_{\gamma'} \sum_{\alpha=1}^{N_p} \dot{\gamma}_p^\alpha \, \boldsymbol{m}_{0p}^\alpha \otimes \boldsymbol{n}_{0p}^\alpha \qquad (14)$$

where, $\dot{\gamma}_j^\alpha$ represents the inelastic shearing rate on slip system $\alpha$, and $\boldsymbol{m}_{0j}^\alpha$ and $\boldsymbol{n}_{0j}^\alpha$ are unit vectors along the slip and slip plane normal, respectively, in the intermediate isoclinic configuration, and $N_j$ denotes the total number of slip systems of the $j^{th}$ phase ($j \in (m, p)$). The inelastic shearing rate, $\dot{\gamma}_j^\alpha$, is a function of the resolved shear stress, $\tau^\alpha$, and internal state variables (ISVs) on each slip system $\alpha$. The slip system-level dislocation density, $\rho_j^\alpha$, is given as the sum of the mobile, $\rho_{Mj}^\alpha$, and immobile, $\rho_{Ij}^\alpha$, dislocation densities, $\rho_j^\alpha = \rho_{Mj}^\alpha + \rho_{Ij}^\alpha$, and is used as an ISV in our framework.

Before moving further, we would like to point out that an iso-stress formulation (or its variant) has also been used in the works of (Estrada Rodas and Neu, 2018; Ma et al., 2008; Tinga et al., 2010) to model the deformation of Ni-based superalloys. Physically, the state of stress is certainly not expected to be the same in both the phases. However, this Sachs-type homogenization provides a lower bound estimate of the deformation and may serve as a first order approximation. Note that this work focuses primarily on predicting the yield stress of two phase Ni-based superalloy single crystals. Given that the respective elastic constants of these two phases only differ by $\approx 10\%$ (cf. table 3), it may be expected that the stresses in the "relatively" coherent phases are not significantly different in the elastic region. Subsequently, the phase with the lower yield stress may start deforming via inelastic deformation and there is expected to be relatively small error in the prediction of the yield stress with the iso-stress assumption; stress/strain concentrations in the two phases may be significant only after yield. Moreover, similar approaches have been extensively employed to model twinning in crystallographic materials (Abdolvand et al., 2011; Kalidindi, 1998). Explicit consideration of the $\gamma$ and $\gamma'$ phases may be employed (Keshavarz and Ghosh, 2013), for example) to obtain higher resolution of the stress and strain states in the individual phases, *albeit* at the cost of significant computational costs. Another approach could be to use a modified Sachs type interaction at the bicrystal interfaces (Tinga et al. (2008, 2010)) to better account for the interfacial deformation. Such approaches may be explored in future work.

*4.2. Crystal plasticity kinetics*

As mentioned earlier, we have 24 octahedral {111}<110> slip systems (considering forward and backward directions) and 6 cube {100}<110> slip systems in the precipitate phase. Further, 12 octahedral {111}<110> slip systems are also present in the fcc matrix phase. Note that these 12 slip systems are of the same family as the octahedral systems in the precipitate phase, without explicit consideration for forward or backward slip. This is purely motivated by computational efficiency considerations of the numerical integration procedure. We have used a fully implicit numerical integration scheme adapted from (Ling et al., 2005) in our framework, the computation time for which increases with the number of slip systems considered. For the precipitate as well as the matrix phase, the first 12 slip systems are the same. For the precipitate phase, slip systems 13 through 24 are the ones with negative directions of the first 12 slip systems. These slip systems are given in table A1 and A2 in the appendix. From here on, we use $\alpha \in oct, j \in m$ to denote octahedral slip in the matrix phase, $\alpha \in oct, j \in p$ to denote octahedral slip in the precipitate phase, and $\alpha \in cube, j \in p$ to denote cube slip in the precipitate phase, respectively. As discussed in the previous sections, only the octahedral slip systems in the precipitate phase demonstrate non-Schmid yield behavior.

An Arrhenius type activation energy based model may be used for modeling thermally-activated dislocation glide in crystalline materials (Kocks et al., 1975). We have used a similar flow rule that gives the crystallographic shearing rate as a function of the driving force due to applied stress and thermal activation due to finite temperature, i.e.,

$$\dot{\gamma}_j^\alpha = \begin{cases} \dot{\gamma}_{0j} exp\left(-\frac{\Delta F_j^\alpha}{kT}\left(1-\left(\frac{|\tau_j^\alpha|-s_{aj}^\alpha}{s_{tj}^\alpha}\right)^{p_j^\alpha}\right)^{q_j^\alpha}\right) sgn(\tau_j^\alpha); \; |\tau_j^\alpha| > s_{aj}^\alpha; \; \alpha \in oct, j \in m \\ \dot{\gamma}_{0j} exp\left(-\frac{\Delta F_j^\alpha}{kT}\left(1-\left(\frac{\tau_j^\alpha-s_{aj}^\alpha}{s_{tj}^\alpha}\right)^{p_j^\alpha}\right)^{q_j^\alpha}\right); \; \tau_j^\alpha > s_{aj}^\alpha; \; \alpha \in oct, j \in p \\ \dot{\gamma}_{0j} exp\left(-\frac{\Delta F_j^\alpha}{kT}\left(1-\left(\frac{|\tau_j^\alpha|-s_{aj}^\alpha}{s_{tj}^\alpha}\right)^{p_j^\alpha}\right)^{q_j^\alpha}\right) sgn(\tau_j^\alpha); \; |\tau_j^\alpha| > s_{aj}^\alpha; \; \alpha \in cube, j \in p \\ 0; \; otherwise \end{cases}$$

(15)

Here, $\dot{\gamma}_{0j}$ is the pre-exponential factor, $\Delta F_j^\alpha$ is the activation energy for dislocation glide in the absence of external stress, $k$ is the Boltzmann constant, $T$ is the absolute temperature, $\tau_j^\alpha$ is the driving stress for dislocation glide, $s_{aj}^\alpha$ is the athermal slip resistance, $s_{tj}^\alpha$ is the thermal resistance to dislocation glide, and $p_j^\alpha$ and $q_j^\alpha$ are parameters used to model the shape of the activation enthalpy function. Note that the signum function (represented by $sgn$) of $\tau_j^\alpha$ accounts for forward and backward slip on the slip system $\alpha$ for octahedral slip in the matrix phase and cube slip in the precipitate phase.

Note that other forms of flow rule have also been used in the literature. For example, power law models have been used in crystal plasticity and dislocation dynamics frameworks (Staroselsky and Cassenti, 2010; Wu et al., 2018) and sine hyperbolic models have been used by (Keshavarz et al., 2016; Roters et al., 2010) for crystal plasticity frameworks. In such flow rules, the temperature dependence is accounted for in the pre-exponential term, generally representative of dislocation mobility. Further, (Kocks et al., 1975) have also provided a detailed analysis of the applicability of the phenomenological thermally-activated flow rule (the one used in our work) to different scenarios by changing the shape function parameters. It was shown that using certain shape function parameter values, this flow rule may even be reduced to a power law form. Finally, it should also be noted that the thermally-activated flow rule has been used extensively (Kothari and Anand, 1998; Ma et al., 2007; Thool et al., 2020) to model the thermo-mechanical deformation of metallic systems.

*4.3. Driving stress for dislocation glide*

The driving stress, $\tau_j^\alpha$, on each slip system is thus a function of the resolved shear stresses, both Schmid and non-Schmid, depending on the type of slip system. Additionally, the non-Schmid stresses, $\tau_{pe}^\alpha$, $\tau_{se}^\alpha$ and $\tau_{cb}^\alpha$ (cf. Section 2), influence deformation on the octahedral systems in the precipitate phase. As discussed in (Patra et al., 2014), the contribution of non-Schmid stresses to the driving stress is expected to decrease with increasing inelastic deformation. This is because past the point of initial yield, dislocation glide is driven primarily by the Schmid stress. Physically, the dislocation core, once constricted at yield, is expected to lie on a single plane during glide. Following this earlier work (Patra et al., 2014), the contribution of non-Schmid stresses to the driving stress are assumed to decay in an exponential fashion. Accordingly, the driving stress, $\tau_j^\alpha$, for the different slip systems is given as

$$\tau_j^\alpha = \begin{cases} \boldsymbol{m}^\alpha \cdot \boldsymbol{\sigma} \cdot \boldsymbol{n}^\alpha; \ \alpha \in oct, j \in m \\ \boldsymbol{m}^\alpha \cdot \boldsymbol{\sigma} \cdot \boldsymbol{n}^\alpha + \left(a_1 \boldsymbol{m}^{pe} \cdot \boldsymbol{\sigma} \cdot \boldsymbol{n}^{pe} + a_2 \boldsymbol{m}^{se} \cdot \boldsymbol{\sigma} \cdot \boldsymbol{n}^{se} + a_3 \boldsymbol{m}^{cb} \cdot \boldsymbol{\sigma} \cdot \boldsymbol{n}^{cb}\right) exp\left(-\frac{\varepsilon_p^i}{\varepsilon_0^i}\right); \ \alpha \in oct, j \in p \\ \boldsymbol{m}^\alpha \cdot \boldsymbol{\sigma} \cdot \boldsymbol{n}^\alpha; \ \alpha \in cube, j \in p \end{cases}$$

(16)

Here, $\varepsilon_p^i$ is the effective inelastic strain in the precipitate phase and $\varepsilon_0^i$ is a material constant. The physicality of the exponential decay model for non-Schmid stresses lies in the fact that the inelastic flow due to dislocation glide is generally expected to be driven solely by the resolved shear stress on the primary slip plane, without contribution from non-Schmid stresses. A parametric study has been performed in section 5.5 to explore the effect of the exponential decay model on the flow stress of the material.

Before moving further, we note that there is no explicit consideration of misfit stresses in the present work. Misfit stresses are generally present at the interfaces of these coherent phases in Ni-based superalloys and are expected play a role during fatigue, creep and also the rafting phenomena observed at higher temperatures (Fan et al., 2015; Grant et al., 2011; Link et al., 2000). Effect of these misfit stresses are assumed to reflect only on the slip resistances, *albeit* indirectly, as described below.

*4.4. Slip resistances*

The athermal slip resistance, $s_a^\alpha$, represents the resistance to dislocation glide due to the long-range stress fields present at the level of grains. Depending on the phase and the type of slip system, there may be multiple dominant contributions to this: (a) grain boundaries and sub-grain structures (Hall-Petch effect), (b) Orowan hardening due to bowing of dislocations around the precipitates, and (c) existing dislocations (Taylor hardening). Accordingly, the athermal slip resistance is modeled as

$$s_{aj}^\alpha = \begin{cases} \tau_{0j}^\alpha + k_{oro}\frac{G_j b_j}{l_{ch}} + \frac{k_{Hj}}{\sqrt{d_\gamma}} + k_{\rho j} G_j b_j \sqrt{\sum_{\zeta=1}^{N_m} A^{\alpha\zeta} \rho_j^\zeta} \ ; \ \alpha \in oct, j \in m \\ \tau_{0j}^\alpha + \frac{k_{Hj}}{\sqrt{d_{\gamma'}^{eff}}} + k_{\rho j} G_j b_j \sqrt{\sum_{\zeta=1}^{N_p} A^{\alpha\zeta} \rho_j^\zeta} \ ; \ \alpha \in (oct, cube), j \in p \end{cases}$$

(17)

where, $\tau_{0j}^\alpha$ is a material constant, $k_{Hj}$ is the Hall-Petch coefficient, $d_\gamma$ is the mean grain size of the $\gamma$ phase, $d_{\gamma'}^{eff}$ is the mean precipitate size of the $\gamma'$ phase, $G_j$ is the shear modulus, $b_j$ is the Burgers vector magnitude, $k_{\rho j}$ is the dislocation barrier strength, and, $A^{\alpha \zeta}$ is the matrix of slip system dislocation interaction coefficients. For a material with large grains ($\sim 1000$ μm) and relative low dislocation density (physically representative of annealed crystals with negligible Hall-Petch effect and/or dislocation hardening), $\tau_{0j}^{\alpha}$ may phenomenologically represent the critically resolved shear stress (CRSS) of the material. This implies that we may directly use the value of $\tau_0^{\alpha}$ that we have estimated in section 3 for the precipitate phase in our crystal plasticity model.

There is an additional term due to Orowan hardening in the matrix phase. This represents the stress required for bowing of dislocations in the narrow matrix channels between precipitates (Kozar et al., 2009; Tinga et al., 2010). This is modeled using $k_{oro}\, G_m b_m/l_{ch}$ in eq. (17), where $k_{oro}$ is a material constant and $l_{ch}$ is the width of the matrix channel.

Finally, the thermal slip resistance, $s_{tj}^{\alpha}$, is a short range stress that may be overcome by lattice thermal vibrations.

As discussed earlier, misfit stresses at the interfaces of the two constituent phases have been neglected in this homogenized formulation. Misfit stresses may either increase or decrease the driving stress in the respective phases (cf. (Tinga et al., 2010)). Further, the misfit stresses may also vary spatially in these single crystal materials (Völkl et al., 1998; Wu and Sandfeld, 2016). These considerations may give rise to a "back stress" in the vicinity of the γ/γ' interfaces and has been successfully modeled in spatially resolved discrete dislocation dynamics simulations (Huang et al., 2012; Wu et al., 2017). However, explicit consideration for these misfit-induced back stresses may perhaps be more difficult in our homogenized formulation. Further, length scale-driven diffusion stress terms (Groma et al., 2016; Wu et al., 2018) may also play a role in these heterogeneously deforming microstructures. As a first order approximation, the $\tau_{0j}^{\alpha}$ term may be assumed to phenomenologically represent any differential stresses experienced by the two phases in the present framework. We leave explicit consideration of these misfit stresses to future work.

*4.5. Substructure evolution*

Constitutive equations governing the evolution of mobile and immobile dislocations have been adapted from (Patra et al., 2014). The rate of evolution of the mobile and immobile dislocations is given as

$$\dot{\rho}_{Mj}^{\alpha} = \begin{cases} \frac{k_{mj}^{\alpha}}{b_j}\left(\sqrt{\sum_{\zeta}\rho_j^{\zeta}} + \frac{1}{d_{\gamma\prime}^{eff}}\right)|\dot{\gamma}_j^{\alpha}| - \frac{2R_c}{b_j}\rho_{Mj}^{\alpha}|\dot{\gamma}_j^{\alpha}| - \frac{k_{tj}^{\alpha}}{b_j\lambda_j^{\alpha}}|\dot{\gamma}_j^{\alpha}|; \alpha \in oct, j \in m \\ \frac{k_{mj}^{\alpha}}{b_j}\sqrt{\sum_{\zeta}\rho_j^{\zeta}}|\dot{\gamma}_j^{\alpha}| - \frac{2R_c}{b_j}\rho_{Mj}^{\alpha}|\dot{\gamma}_j^{\alpha}| - \frac{k_{tj}^{\alpha}}{b_j\lambda_j^{\alpha}}|\dot{\gamma}_j^{\alpha}|; \alpha \in (oct, cube), j \in p \end{cases} \quad (18)$$

$$\dot{\rho}_{Ij}^{\alpha} = \frac{k_{tj}^{\alpha}}{b_j\lambda_j^{\alpha}}|\dot{\gamma}_j^{\alpha}| - k_{dj}^{\alpha}\rho_{Ij}^{\alpha}|\dot{\gamma}_j^{\alpha}|; \alpha \in (oct, cube), j \in (m, p) \quad (19)$$

The first term on the RHS of eq. (18) models the multiplication of mobile dislocations at pre-existing dislocation segments (Essmann and Mughrabi, 1979), with $k_{mj}^{\alpha}$ being the dislocation multiplication constant. The second term models the annihilation of dislocation dipoles within a capture radius, $R_c$. The third term models the trapping of mobile dislocation segments at other dislocations. The effective free path of dislocation trapping is a function of the mean free path of dislocations:

$$\lambda_j^{\alpha} = \begin{cases} 1/\sqrt{\rho_j^{\alpha}} + d_{\gamma\prime}^{eff}; \alpha \in oct, j \in m \\ 1/\sqrt{\rho_j^{\alpha}}; \alpha \in (oct, cube), j \in p \end{cases} \quad (20)$$

$k_{tj}^{\alpha}$ is the dislocation trapping constant. As a result of trapping, the mobile dislocations get immobilized, as represented by the first term on the right hand side of eq. (19). These immobile dislocations may be annihilated by dynamic recovery (represented using the second term on the right hand side of eq. (19)) and $k_{dj}^{\alpha}$ is the dynamic recovery constant. The material constants, $k_{mj}^{\alpha}$, $R_c$, $k_{tj}^{\alpha}$ and $k_{dj}^{\alpha}$, are to be determined by fitting the constitutive model to the experimental data. Note that the dislocation multiplication term and the trapping term for the matrix phase in eq. (18) account for the effective spacing between the precipitates. As a first order approximation, these are assumed to be equal to the mean precipitate size, $d_{\gamma\prime}^{eff}$, calculated in eq. (12).

These set of equations are assumed to represent the dominant mechanisms governing the quasi-static deformation of Ni-based superalloys. Model considerations for anisotropic creep deformation of Ni-based superalloys may be incorporated in future work. The framework may also be applied for life assessment of gas turbine blades and structures (Narasimhachary et al., 2019).

These constitutive equations have been implemented as a material model and interfaced with the open source finite element framework, Multiphysics Object Oriented Simulation Environment (MOOSE) (Gaston et al., 2009).

*4.6. Model parameters*

Calibration of the parameters for Ni-based superalloys is rather challenging given the number of physical mechanisms and model parameters involved. We have adopted a modular strategy for accomplishing this. We first calibrate the model parameters to the experimental data for Ni$_3$Al single crystals (with $f_{\gamma'} = 1$). Subsequently, we estimate model parameters for the $\gamma$ phase of CMSX-4 based on fit to the experimental data for the two phase alloy.

The temperature dependent elastic constants for the $\gamma$ and $\gamma'$ phases are given in table 3.

**Table 3.** Temperature-dependent elastic constants for the $\gamma$ and $\gamma'$ phases (from (Keshavarz et al., 2016)). All temperatures are in absolute units.

| Phase | Value |
|---|---|
| $\gamma$ | $C_{11} = (298.0 - 0.096T)$ GPa <br> $C_{12} = (191.0 - 0.057T)$ GPa <br> $C_{44} = (139.0 - 0.035T)$ GPa |
| $\gamma'$/Ni$_3$Al | $C_{11} = (325.0 - 0.096T)$ GPa <br> $C_{12} = (209.0 - 0.057T)$ GPa <br> $C_{44} = (144.0 - 0.035T)$ GPa |

As discussed in the preceding sections, the main mechanism governing the inelastic deformation of Ni$_3$Al is due to the formation of KW locks, which manifests in terms of the non-Schmid yield behavior. The initial CRSS ($\tau_{0,oct}^{\alpha}$) for Ni$_3$Al was estimated in section 2. This is directly used in the model for athermal slip resistance in eq. (17). Further, the threshold resistance for cube slip, $\tau_{0,cube}^{\alpha}$, is taken from (Allan, 1995). This is given in eq. (11). These are the main parameters influencing the yield behavior of Ni$_3$Al single crystals. These single crystals are assumed to have a very low initial dislocation density of $\rho_M^0 = \rho_I^0 = 1 \times 10^{10} \text{m}^{-2}$. Note that this may be considered a lower bound estimate of the dislocation density in fully annealed crystals. Further, a dislocation barrier strength of $k_{\rho,oct} = k_{\rho,cube} = 0.31$ is assumed.

The Hall-Petch effect is assumed to be inactive during the initial calibration of the model to Ni$_3$Al single crystal experimental data. This mechanism is expected to be dominant only during the deformation of the two phase microstructures with sub-micron sized precipitates. The thermal slip resistance is assumed to have a very small value of 5 MPa on all slip systems. The Burger vector magnitude is taken from (Keshavarz and Ghosh, 2015). The flow parameters in eq. (15) are chosen such that a single set of parameters are valid for all temperatures. We have chosen physically representative values of the dislocation hardening parameters, without calibrating to the actual hardening response. While there is some experimental data available for the post yield hardening behavior of Ni$_3$Al crystals (Aoki and Izumi, 1979; Heredia and Pope, 1991), we restrict ourselves to fitting to the anomalous yield stress only. As reported in (Aoki and Izumi, 1979), the hardening response itself is orientation- and temperature-dependent and may lead to further complexities. Nonetheless, qualitative trends for the post-yield hardening behavior are discussed with respect to relevant experiments in section 5.5.

These material parameters are listed in table 4. It should be noted that the experimental yield stress data are available for over 80 combinations of loading orientations, types and test temperatures. Fitting to this large data set via trial and error or even other optimization methods would have been significantly time consuming and/or computationally intensive. However, we have bypassed this by estimating the initial CRSS values from the experimental data and using them directly in our model. Of course, some additional calibration is required for the flow rule and hardening parameters. As an alternate scenario, if the non-Schmid parameters were to be estimated via trial and error by individually running CPFE simulations for each orientation and temperature, this would potentially require a large number of calibration simulations to be run and significant computational efforts (especially, given the extensive temperature range over which experimental data is reported). However, by analytically estimating the initial CRSS values *a priori*, the calibration effort is hugely reduced and restricted only to the flow and hardening parameters.

After obtaining a reasonable fit to the experimental yield stress data for Ni$_3$Al, we performed calibration of the homogenized two-phase model to the experimental yield stress data for CMSX-4. To the extent possible, the model parameters for the $\gamma'$ phase for CMSX-4 are assumed to be the same as that of Ni$_3$Al. This is under the implicit assumption that, the material chemistry notwithstanding, the $\gamma'$ phase of CMSX-4 has the same mechanical properties (and model parameters) as Ni$_3$Al crystals. However, some modifications have to be made to the non-Schmid coefficients for the $\gamma'$ phase of CMSX-4. An inspection of the experimental yield stress data for CMSX-4 (cf. section 5.3) revealed that there is significant tension-compression asymmetry for the [0 1 1] loading orientation at lower temperatures ($T \leq 673$ K), while no such asymmetry was observed for Ni$_3$Al. Clearly, the non-Schmid parameters fit to the relatively low tension-compression asymmetry of Ni$_3$Al, will not be able to model the observed asymmetry in CMSX-4. In order to account for this, we have modified $a_2$ and $a_3$ according to the following piecewise linear equations:

$$a_2 = \begin{cases} 0; & T \leq 298 \text{ K} \\ 0.1 \frac{(T - 298)}{(673 - 298)}; & 298 \text{ K} < T \leq 673 \text{ K} \\ 0.1; & T > 673 \text{ K} \end{cases} \quad (21)$$

$$a_3 = \begin{cases} -0.18; & T \leq 873 \text{ K} \\ -0.18 + 0.12 \frac{(T - 873)}{(922 - 873)}; & 873 \text{ K} < T \leq 922 \text{ K} \\ -0.06; & T > 922 \text{ K} \end{cases} \quad (22)$$

The non-Schmid coefficients for CMSX-4 are plotted in figure 4. Note that $\tau_{0,oct}^{\alpha}$ remains the same as earlier.

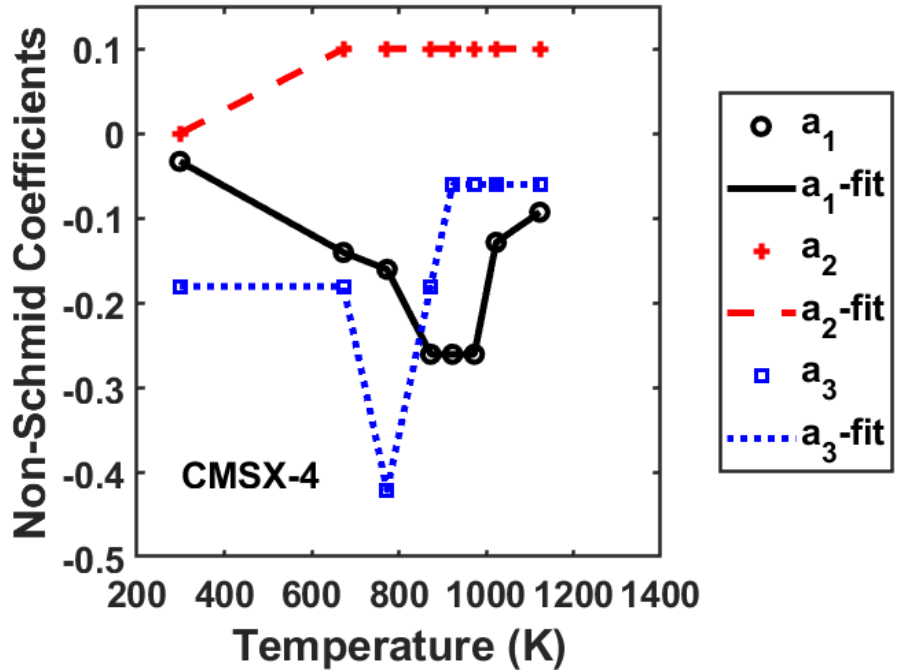

**Figure 4.** Variation of non-Schmid coefficients, $a_1$, $a_2$ and $a_3$ with temperature for octahedral slip in CMSX-4.

We had earlier neglected the Hall-Petch effect for Ni$_3$Al single crystals. However, this is expected to be non-negligible for the $\gamma'$ phase. This is due to the microstructural considerations of the two phase CMSX-4 alloy. The CMSX-4 microstructure is comprised of cuboidal precipitates of size 500 nm, separated by continuous matrix channels of width, $l_{ch} = 60$ nm (Tinga et al., 2010). These microstructural parameters are listed in table 5. At these sub-micron length scales, the Orowan hardening effect in the matrix phase and the Hall-Petch effect in the precipitate phase are expected to play a dominant role. As discussed in (Tinga et al., 2008), the material constant associated with Orowan hardening, $k_{oro}$, takes values between 0.238 and 2.15. We have used a value of $k_{oro} = 0.25$ in our model. The Hall-Petch coefficient has been calibrated as a function of temperature to model the decrease in contribution of the Hall-Petch effect with increase in temperature. This follows from studies for various metals and alloys in the literature (Ono et al., 2004; Schneibel et al., 2011; Zhao et al., 2003), where the contribution of grain size hardening was found to decrease at elevated temperatures. The Hall-Petch coefficients fit to the experimental data are of the following form:

$$k_{Hj}(MPa\sqrt{\mu m}) = \begin{cases} 3.635 \times 10^2 - 7.487 \times 10^{-1}T + 1.087 \times 10^{-3}T^2 - 5.988 \times 10^{-7}T^3; \alpha \in oct, j \in (m,p) \\ \begin{cases} 2.182 \times 10^2 exp(-T/993.442); T \leq 923\ K \\ 3.561 \times 10^2 - 2.839 \times 10^{-1}T; T > 923\ K \end{cases}; \alpha \in cube, j \in p \end{cases}$$

(23)

The dislocation hardening parameters for the $\gamma$ phase are assumed to be similar to that of the $\gamma'$ phase. While the initial dislocation densities are assumed to be the same for the $\gamma'$ phase, the $\gamma$ phase is assumed to have a somewhat higher initial dislocation density of $\rho_M^0 = \rho_I^0 = 1 \times 10^{11} m^{-2}$. This is based on experimental studies (Agudo Jácome et al., 2013; Pollock and Argon, 1992), where dislocation densities of the order $10^{11} - 10^{12} m^{-2}$ have been reported. While phase specific dislocation densities were not reported in these studies, it was noted that dislocations, if any, were mainly observed in the $\gamma$ channels. Finally, the CRSS, $\tau_{0,oct}^\alpha$, of the $\gamma$ phase is modeled as a temperature dependent function to fit to the experimental yield stress data of CMSX-4 according to the following:

$$\tau_{0,j}^\alpha(MPa) = \begin{cases} 3.709 \times 10^2 + 7.095 \times 10^{-2}T - 3.817 \times 10^{-5}T^2; T \leq 923\ K \\ 5.891 \times 10^4 - 1.698 \times 10^2 T + 1.642 \times 10^{-1}T^2 - 5.297 \times 10^{-5}T^3; T > 923\ K \end{cases}; \alpha \in oct, j \in m$$ (24)

The entire set of model parameters is given in table 4. We would like to point out that this set of material parameters is not unique. Other parameter sets may also give model predictions that match the experimental data. In our work, the focus has been to ensure that the values lie within the physically expected range of these parameters, with minimal calibration performed via trial and error.

**Table 4.** Model parameters for the $\gamma$ and $\gamma'$ phases.

| Parameters | $\alpha \in oct, j \in m$ ($\gamma$) | $\alpha \in oct, j \in p$ ($\gamma'$/Ni$_3$Al) | $\alpha \in cube, j \in p$ ($\gamma'$/Ni$_3$Al) |
|---|---|---|---|
| Non-Schmid coefficients: $a_1, a_2, a_3$ | - | Ni$_3$Al: eq. (7), eq. (8), eq. (9) <br> $\gamma'$: eq. (7), eq. (21), eq. (22) | - |
| Non-Schmid decay: $\varepsilon_0^i$ | | Ni$_3$Al: 0.005 <br> $\gamma'$: 0.01 | |
| Flow rule: $\dot{\gamma}_{0j}$ (s$^{-1}$), $\Delta F_j^\alpha, p_j^\alpha, q_j^\alpha$ | $1.0 \times 10^2$, $0.35Gb^3$, 0.3, 1.7 | 1.0, $1.5Gb^3$, 0.3, 1.7 | $1.0 \times 10^{-2}$, $4.5Gb^3$, 1.0, 1.0 |
| Threshold slip resistance: $\tau_{0j}^\alpha$ | eq. (24) | eq. (10) | eq. (11) |
| Hall-Petch coefficient: $k_{Hj}$ | eq. (23) | eq. (23) | eq. (23) |
| Orowan hardening: $k_{oro}$ | 0.25 | - | - |
| Thermal slip resistance: $s_{tj}^\alpha$ (MPa) | 10.0 | 5.0 | 5.0 |
| Dislocation hardening: $k_{\rho j}, A^{\alpha\alpha}, A^{\alpha\zeta} (\alpha \neq \zeta)$ | 0.31, 1.0, $1.0 \times 10^{-4}$ | 0.31, 1.0, $1.0 \times 10^{-4}$ | 0.31, 1.0, $1.0 \times 10^{-4}$ |
| Initial dislocation densities: $\rho_M^0$ (m$^{-2}$), $\rho_I^0$ (m$^{-2}$) | $1 \times 10^{11}$, $1 \times 10^{11}$ | $1 \times 10^{10}$, $1 \times 10^{10}$ | $1 \times 10^{10}$, $1 \times 10^{10}$ |
| Dislocation evolution: $k_{mj}^\alpha, R_c, k_{tj}^\alpha, k_{dj}^\alpha$ | 0.15, $6b$, 0.12, $1.0 \times 10^3$ | 0.07, $6b$, 0.065, $6.0 \times 10^2$ | 0.07, $6b$, 0.065, 6.0 |
| Burgers vector magnitude: $b$ (nm) | 0.249 | 0.249 | 0.249 |

**Table 5.** Microstructural parameters for Ni$_3$Al and CMSX-4. Values for CMSX-4 adapted from (Tinga et al., 2010).

| Material | $\gamma$ phase | $\gamma'$ phase |
|---|---|---|
| Ni$_3$Al | - | $f_{\gamma'} = 1, d_{\gamma'}^{eff} = 1000$ µm |
| CMSX-4 | $f_\gamma = 0.28, d_\gamma = 500$ µm, $l_{ch} = 0.06$ µm | $f_{\gamma'} = 0.72, d_{\gamma'}^{eff} = d_{\gamma'}^p = 0.5$ µm |

## 5. Model predictions and discussion

We first discuss the CPFE model predictions for Ni$_3$Al single crystals and then discuss the model predictions for single crystal Ni-based superalloy, CMSX-4. All predictions have been made using single element calculations. A hexahedral finite element with linear interpolation and full integration is used. Symmetric boundary conditions are applied on three adjacent faces of the simulation cell, meaning that the displacements normal to these faces are set to be zero. The corner node, common to these three faces, is fixed in all degrees of freedom to prevent rigid body motion. Displacement controlled loading (either tensile or compressive) is applied along the z-direction in all cases. These boundary conditions are shown in figure 5(a). A nominal strain rate of $1.3 \times 10^{-3}$/s has been used in the simulations for Ni$_3$Al, while a strain rate of $1.0 \times 10^{-3}$/s has been used in the simulations for CMSX-4 to correspond with the experiments.

*5.1. Effect of orientation and temperature on the yield stress of Ni$_3$Al crystals*

The model predictions are verified against the calibrated yield stress of Ni$_3$Al single crystals for four loading orientations: [0 0 1], [$\bar{1}$ 3.2 16], [$\bar{1}$ 9.1 10.1] and [$\bar{1}$ 1.1 1.2] over the temperature range of $260 - 1304$ K. These are the loading orientations for which the non-Schmid model parameters have been estimated in section 3. These orientations are shown inside the standard stereographic triangle in figure 5(b). [0 0 1], [$\bar{1}$ 9.1 10.1] and [$\bar{1}$ 1.1 1.2] represent loading orientations at/near the three corners of the standard stereographic triangle, while the [$\bar{1}$ 3.2 16] loading orientation lies somewhere in between these orientations. The crystal axes are assumed to be aligned with the laboratory axes (in the Cartesian coordinate system) for [0 0 1] loading, while the crystal is rotated using appropriate Euler angles (to denote the rotation of the single crystal with respect to the loading axis) for the other loading orientations. Comparison of CPFE model predictions of the yield stress with the experimental data (Heredia and Pope, 1991) for these loading orientations are shown in figure 6(a-d).

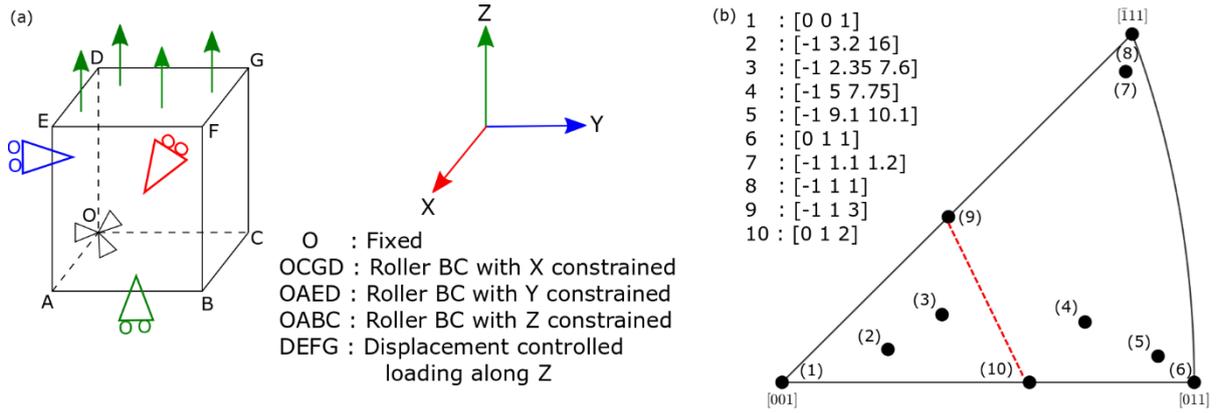

**Figure 5.** (a) Schematic of loading and boundary conditions used in the simulations. (b) Loading orientations for which the yield stress predictions are made in the present work. The red line joining the $[\bar{1}\ 1\ 3]$ and $[0\ 1\ 2]$ orientations indicates the region where zero tension-compression asymmetry is expected.

Note that all model predictions are for the proportional limit stress. While most experimental data are reported for the 0.2% offset yield stress, we choose the stress at proportional limit as the measure of yield stress in our simulations. This is because the proportional limit stress represents the (theoretical) stress at which inelastic deformation commences and more importantly the stress to which our non-Schmid yield model has been calibrated. We have observed that there are non-negligible differences between the proportional limit yield stress and the 0.2% offset yield stress in many of our model predictions. Similar observations have also been made by (Keshavarz and Ghosh, 2015). Such differences may be due to the effects of temperature, strain rate and orientation on the post-yield inelastic flow.

The model predictions do not exhibit any significant difference in yield stress between tension and compression for either of the four orientations at lower temperatures (260 − 300 K). This is in agreement with the experiment trends (Heredia and Pope, 1991). In the temperature range 400 − 1000 K, the model predicts higher yield stress in tension than in compression for the $[0\ 0\ 1]$ (figure 6(a)) and $[\bar{1}\ 3.2\ 16]$ orientations (figure 6(b)), as also observed in experiments. This difference is attributed to the yield being governed primarily by octahedral slip. A simple calculation of the Schmid factors shows that cube slip is not feasible for loading orientations near the $[0\ 0\ 1]$ corner of the standard stereographic triangle. A peak yield stress is obtained at 973 K in tension with a value of 877 MPa in tension and at 1105 K in compression with a value of 588 MPa. The difference in yield stress between tension and compression is maximum in this temperature range, and decreases with further increase in temperature. It is noted that there is some discrepancy in the peak value of the yield stress between model predictions and experiments, especially for the [$\bar{1}$ 3.2 16] loading orientation in tension (figure 6(b)). Possible reasons are described at the end of this section.

The yield behavior for the [$\bar{1}$ 9.1 10.1] (figure 6(c)) and [$\bar{1}$ 1.1 1.2] (figure 6(d)) loading orientations show an opposite trend as compared to the [0 0 1] and [$\bar{1}$ 3.2 16] orientations in the temperature range of $400 - 900$ K: the predicted yield stress is higher in compression than in tension. Yield is governed by octahedral slip up to $\approx 900$ K for [$\bar{1}$ 1.1 1.2] loading and $\approx 1100$ K for [$\bar{1}$ 9.1 10.1] loading. Based on geometric and non-Schmid considerations, we expect to have a higher yield stress in compression for these orientations. Cube slip gets activated beyond these temperatures, respectively, for each of the above orientations. This explains the absence of tension-compression asymmetry for these loading orientations at higher temperatures; non-Schmid effects do not contribute to cube slip. Overall, the model predictions compare reasonably with experiments and this verifies the validity of our approach for estimating the non-Schmid parameters.

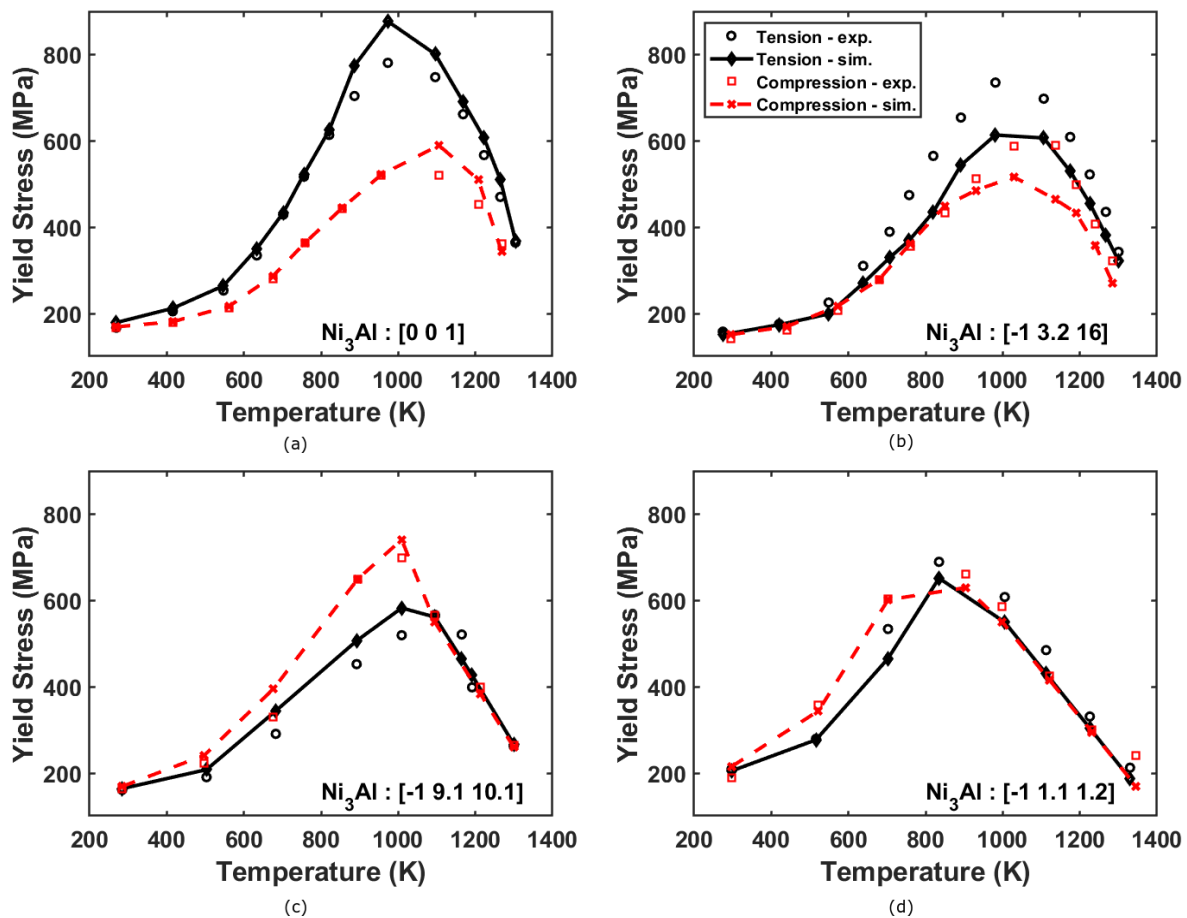

**Figure 6.** Comparison of CPFE model predictions of the temperature-dependent yield stress of Ni$_3$Al single crystals with the experimental data for (a): [0 0 1], (b): [$\bar{1}$ 3.2 16], (c): [$\bar{1}$ 9.1 10.1], and (d): [$\bar{1}$ 1.1 1.2] loading orientations. The experimental data (in open symbols) is adapted from (Heredia and Pope, 1991).

We have also used the model to predict the yield stress for two loading orientations for which the non-Schmid model parameters have not been calibrated. Figure 7 shows CPFE model predictions of the yield stress as a function of temperature for the [$\bar{1}$ 2.35 7.6] and [$\bar{1}$ 5 7.75] loading orientations, both in tension and compression. These orientations are also shown in figure 5(b).

Loading orientation [$\bar{1}$ 2.35 7.6] lies close to the line joining [0 1 2] and [$\bar{1}$ 1 3], and the difference in yield stress between tension and compression is expected to be small over all temperatures for this loading orientation (as $\tau_{pe}$ and $\tau_{se}$ are generally low in the vicinity of this line) (Allan, 1995). Model predictions are compared with the experimental data in figure 7(a). Similar to the experiments, the model predicts minimal difference in the yield stress values between tension and compression over the temperature range of $260 - 1304$ K. The model however fails to predict the temperature of the peak yield stress accurately. While the observed peak in the experimental yield stress data is reached at $\approx 1100$ K, the model predicts the same at lower temperatures.

Loading orientation [$\bar{1}$ 5 7.75] lies on the other side of the line joining [0 1 2] and [$\bar{1}$ 1 3], towards the [0 1 1] and [$\bar{1}$ 1 1] poles of the stereographic triangle. From geometric and non-Schmid considerations, this implies a greater value of yield stress in compression than in tension. Figure 7b shows model predictions of the yield stress with temperature for this loading orientation and compared with the experimental data. At lower temperatures, the difference in yield stress between tension and compression is negligible, but with further increase in temperature, we observed higher value of yield stress in compression than in tension. The yield stress value in compression reaches a peak value of 625 MPa at 924 K, which is in agreement with the experimental data.

A source of discrepancy between the model predictions and experimental yield stress is due to the fact that while the model results are for the proportional limit stress, the experimental values are for the 0.2% offset yield stress. This is also apparent when the predicted proportional limit yield stress and the 0.2% offset yield stress are compared with the experimental data (see figure A4 in the appendix for a comparison of the same orientations as in figure 6). There are non-negligible differences in many cases. As mentioned earlier, the choice of this criterion is because the non-Schmid model has been calibrated to predict the point at which inelastic deformation commences. Macroscopically, this manifests as a deviation from linearity in the stress-strain response and is generally known as the proportional limit stress. Beyond the proportional limit stress, the orientation-dependent hardening behavior dictates whether the 0.2% yield stress is significantly different from the proportional limit stress. Further, the effects of strain rate and temperature may also manifest in the overshoot of the viscous stress beyond this point, particularly for materials that have a high rate sensitivity. These are the potential reasons that model predictions are not in quantitative agreement with the experimental values in certain cases. Quantitative discrepancies notwithstanding, the model predictions and trends are in general agreement with the experiments. Since the model is generally able to predict the experimental data, including orientations for which the non-Schmid parameters have not been calibrated, this serves as an overall validation of the modeling approach to predict orientation- and temperature-dependent yield stress trends. This is especially important for establishing the model's predictive capability given the fact we have chosen a set of non-Schmid parameters from multiple possible sets of non-unique solutions.

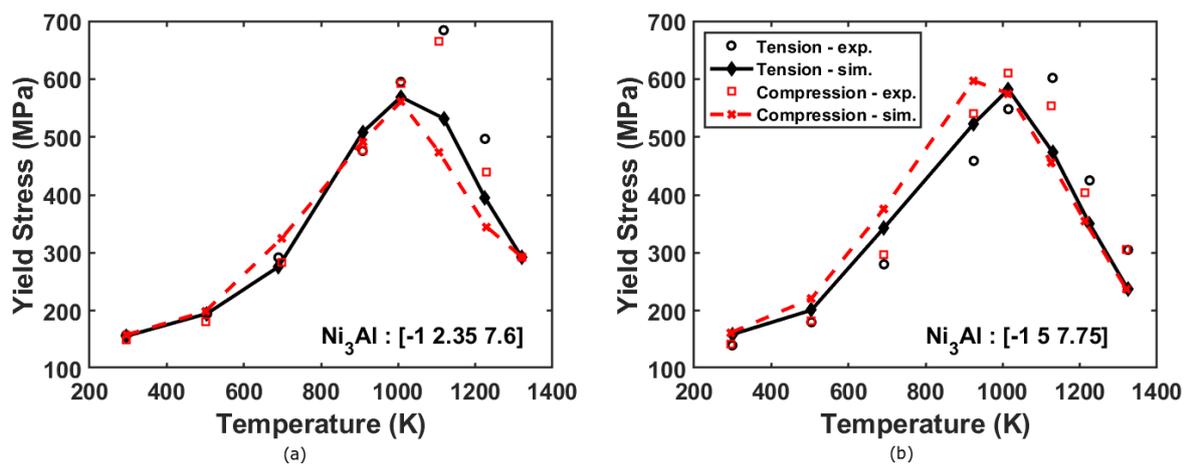

**Figure 7.** Comparison of CPFE model predictions of the temperature-dependent yield stress of $Ni_3Al$ single crystals with the experimental data for (a): $[\bar{1}\ 2.35\ 7.6]$, (b) $[\bar{1}\ 5\ 7.75]$ loading orientations. The experimental data (in open symbols) is adapted from (Heredia and Pope, 1991).

*5.2. Tension-compression asymmetry and slip system activity of Ni$_3$Al*

The deviation from Schmid law in L1$_2$ single crystals is inferred to be the reason for yield anomalies observed and the resultant tension-compression asymmetry. As reported by (Heredia and Pope, 1991), there are different regions in the stereographic triangle which exhibit different degrees of tension-compression asymmetry. Also, there is a region where no asymmetry is expected (Heredia and Pope, 1991). Although this has been confirmed by experiments and simulations for some specific loading orientations by our CPFE model predictions and also in the works of (Allan, 1995; Keshavarz and Ghosh, 2015), we try to explore these anomalies observed over the entire standard stereographic triangle at different temperatures.

We study tension-compression asymmetry using the strength differential parameter, *SD*, (Bassani and Racherla, 2011):

$$SD = \frac{(\sigma_Y^T - \sigma_Y^C)}{(\sigma_Y^T + \sigma_Y^C)/2} \qquad (25)$$

where, $\sigma_Y^T$ is the yield stress in tension and $\sigma_Y^C$ is the yield stress in compression. A positive value of *SD* indicates higher yield stress in tension, while a negative value is indicative of higher yield stress in compression.

We have analytically computed the *SD*, with the orientation-dependent Schmid factors and the CRSS values reported earlier, using an in-house MATLAB code. Figure 8 shows the *SD* contours of Ni$_3$Al crystals plotted over the standard stereographic triangle for different temperatures. The dominant (octahedral or cube) slip activity at yield is also plotted over the stereographic triangle for the crystal loaded in tension and in compression in figure 8. In the slip system activity contours, blue color indicates dominant octahedral slip at yield, while red color indicates dominant cube slip at yield.

There is a mild tension-compression asymmetry observed at 300 K for all orientations. This is in coherence with the experimental data (Heredia and Pope, 1991), according to which no tension-compression asymmetry is observed at lower temperatures $270 - 300$ K. The analytical predictions indicate significant asymmetry at 600 K. As observed earlier in the yield stress data, orientations near the [0 0 1] pole attain higher value of yield stress in tension than in compression, while orientations near the [0 1 1] and [$\bar{1}$ 1 1] poles show greater yield stress in compression. The slip system activity plots for 300 K and 600 K, for both tension and compression, indicate that yield is governed by the octahedral slip over the entire stereographic triangle. This result also is in coherence with the observations in literature (Heredia and Pope, 1991).

The asymmetry becomes more pronounced with increase in temperature. At 800 K and 900 K, a positive $SD$ is observed for loading orientations near the [0 0 1] pole, while a negative $SD$ is observed for loading orientations near the [0 1 1] and [$\bar{1}$ 1 1] poles. This can be attributed to the dominance of octahedral slip and ensuing KW lock formation (cf. section 2). At 800 K and 900 K, orientations near the [0 1 1] and [$\bar{1}$ 1 1] poles, especially those near the [$\bar{1}$ 1 1] pole, show a decrease in the $SD$. The slip activity plots in the range of 800 − 900 K indicate that yielding in tension is still governed by the activation of the octahedral slip, while there is activation of cube slip in compression for these loading orientations. This is in agreement with the experimental data, which suggests the onset of cube slip near the [$\bar{1}$ 1 1] pole at ≈ 873 K in compression (Heredia and Pope, 1991). In a recent experimental study (Coupeau et al., 2020), evidence of cube slip in Ni$_3$Al crystals was reported, while also noting that the KW locks formed on these planes may be mobile.

At 1000 K, a higher $SD$ value for loading orientations near the [0 0 1] pole may be attributed to the peak value of CRSS for octahedral slip reached at 973 K. This reflects in the slip system activity plots at 1000 K, which show that octahedral slip is active at yield, both in tension and compression. Experimental observations also suggest a higher asymmetry at this temperature (Heredia and Pope, 1991). For orientations near the [$\bar{1}$ 1 1] pole, lower $SD$ indicates that cube slip governs the yield at 1000 K in both tension and compression, as is also confirmed from the slip activity plots.

$SD$ contours for 1200 K indicate that the tension-compression asymmetry is generally lower. For orientations close to the [0 0 1] pole, the deformation is still dominated by slip on the octahedral slip system and hence a small positive value of $SD$. But for orientations close to [0 1 1] and [$\bar{1}$ 1 1] poles, deformation is now dominated by cube slip, and thus there is no $SD$.

The region joining the [0 1 2] and [$\bar{1}$ 1 3] orientations in the standard stereographic triangle in figure 5 exhibits no tension-compression asymmetry for all temperatures. This is due to the fact that the combined effect of non-Schmid stresses, $\tau_{pe}$ and $\tau_{se}$, is zero for loading orientation in this region. At temperatures above 1000 K, the region on the left side of the line joining [0 1 2] and [$\bar{1}$ 1 3] orientations (towards the [0 0 1] pole) experiences the activation of octahedral slip, while the region to the right of this line experiences the activation of cube slip in both tension and compression. As discussed earlier, octahedral slip with a small $SD$ is expected for the former, while no $SD$ is expected for the latter region.

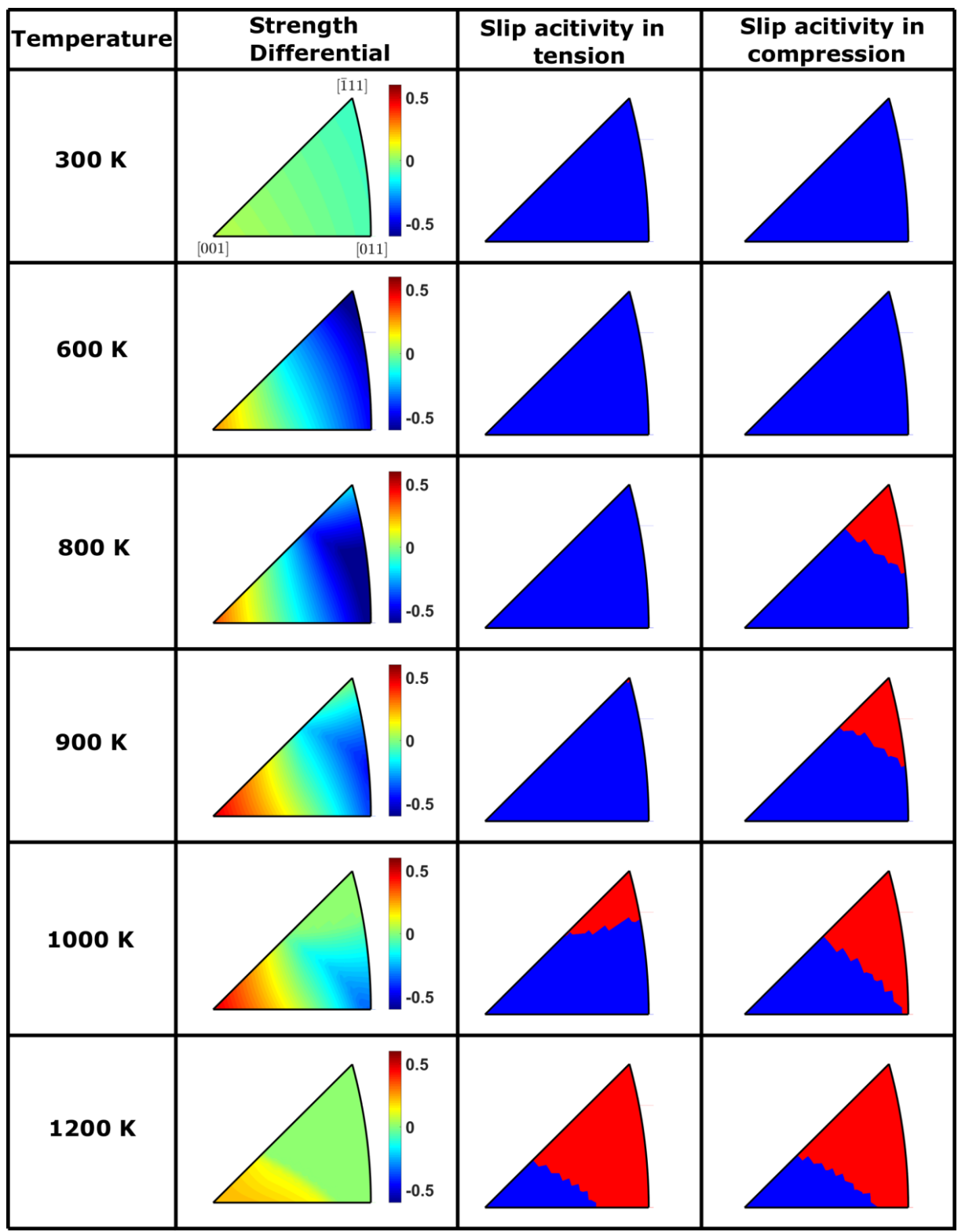

**Figure 8.** Prediction of strength differential (*SD*), and dominant slip activity in tension and compression for Ni$_3$Al. In the slip activity plots, blue color indicates octahedral slip and red color indicates cube slip.

*5.3. Effect of orientation and temperature on the deformation of CMSX-4*

Figure 9 shows the comparison of CPFE model predictions with the experimental yield stress as a function of temperature for CMSX-4 single crystals loaded along three different orientations: [0 0 1], [0 1 1] and [$\bar{1}$ 1 1]. The experimental data is taken from (Allan, 1995).

Generally speaking, model predictions are in good agreement with the experiments for all temperatures and orientations. Moreover, some of the yield stress trends of CMSX-4 are similar to those observed earlier for Ni$_3$Al crystals. For brevity, we do not go into the details of those trends here. Rather, we highlight the notable features and differences between the two materials here.

The most striking difference between Ni$_3$Al and CMSX-4 is regarding the actual values of the yield stress. The yield stress is, in general, higher for CMSX-4 as compared to Ni$_3$Al for all orientations and temperatures. As discussed earlier, this is due to multiple mechanisms: Orowan hardening in the matrix and Hall-Petch effect in the precipitate phase. At lower temperatures ($< 873$ K), the yield stress is generally higher for [0 1 1] and [$\bar{1}$ 1 1] loading orientations, as compared to the [0 0 1] loading orientation. The reverse trend is observed at higher temperatures ($T > 973$ K). Generally speaking, the yield stress does not change significantly up to about 973 K, beyond which it decreases rapidly (with temperature). This plateau in the yield stress behavior (with respect to temperature), observed in both model predictions and experiments, is typical of Ni-based superalloys and allows their use in high temperature gas turbine engine environments (Caron and Khan, 1999). The yield stress is higher in compression than in tension for the [0 1 1] orientation across all temperatures, while the reverse trend is observed for the [0 0 1] orientation. The difference in yield stress between tension and compression is relatively low for the [$\bar{1}$ 1 1] loading orientation. As mentioned earlier, the non-Schmid model parameters, $a_2$ and $a_3$, had to be modified for the $\gamma'$ phase in order to account for this observed asymmetry in the yield behavior at lower temperatures.

The orientation- and temperature-dependent yield of CMSX-4 is further complicated due to the availability of another slip mode, namely, octahedral slip in the $\gamma$ phase, in addition to octahedral and cube slip in the $\gamma'$ phase. We have summarized the activation of different slip systems at yield for the various loading conditions in table 6. We discuss this data in the next section in the context of the slip system activity contours and the $SD$ contours.

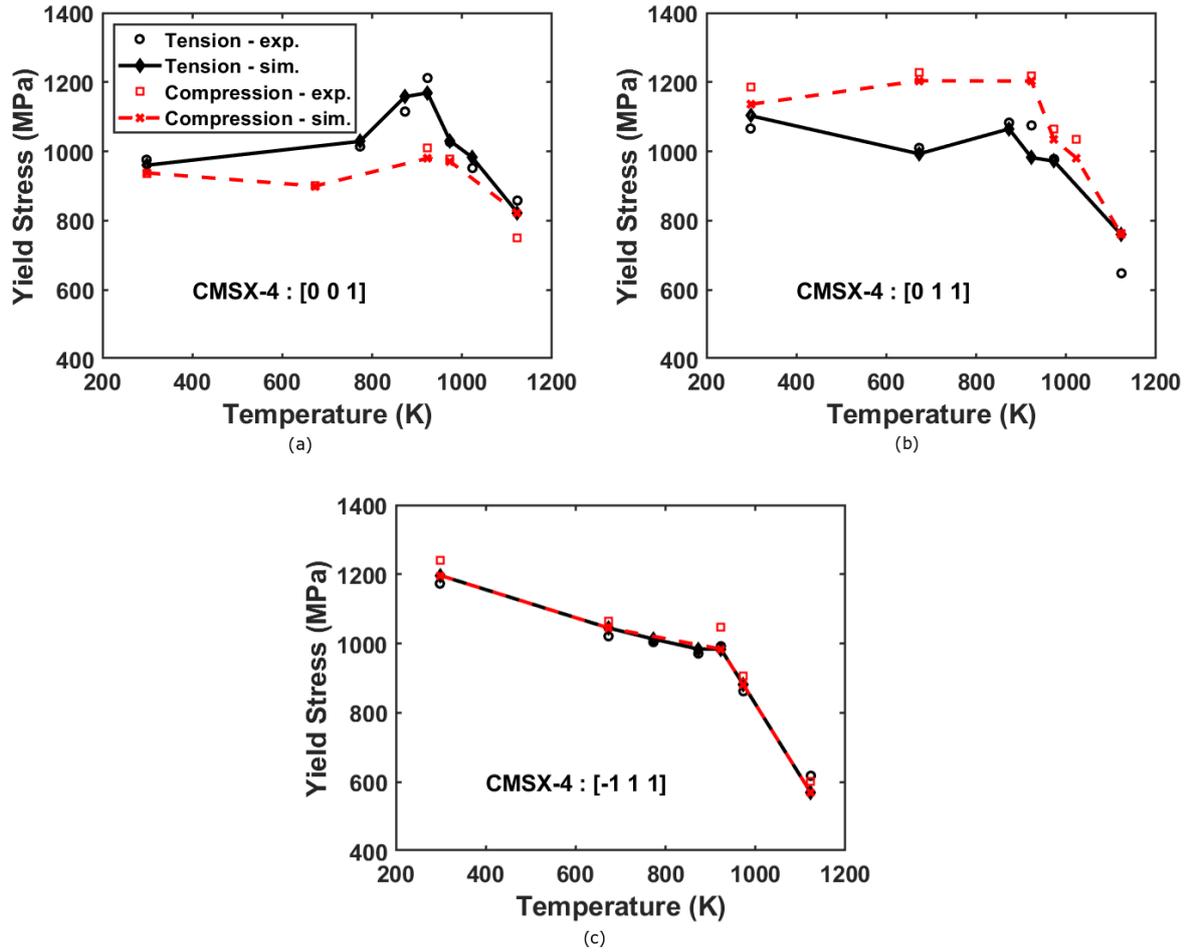

**Figure 9.** Comparison of CPFE model predictions of the temperature-dependent yield stress of CMSX-4 single crystals with the experimental data for (a): [0 0 1], (b): [0 1 1], and (c): [$\bar{1}$ 1 1] loading orientations. The experimental data is adapted from (Allan, 1995).

**Table 6.** Dominant slip modes in CMSX-4 at yield for the various loading conditions. T indicates tension and C indicates compression.

| Temperature (K) | Loading Orientation/Type | | | | | |
|---|---|---|---|---|---|---|
| | [0 0 1]/T | [0 0 1]/C | [0 1 1]/T | [0 1 1]/C | [1 1 1]/T | [1 1 1]/C |
| 298 | $\gamma'$, octahedral | $\gamma'$, octahedral | $\gamma'$, octahedral | $\gamma'$, octahedral | $\gamma'$, cube | $\gamma'$, cube |
| 673 | $\gamma'$, octahedral | $\gamma'$, octahedral | $\gamma'$, octahedral | $\gamma'$, octahedral | $\gamma'$, cube | $\gamma'$, cube |
| 773 | $\gamma'$, octahedral | $\gamma'$, octahedral | $\gamma'$, octahedral | $\gamma'$, octahedral | $\gamma'$, cube | $\gamma'$, cube |
| 873 | $\gamma'$, octahedral | $\gamma'$, octahedral | $\gamma'$, octahedral | $\gamma'$, octahedral | $\gamma'$, cube | $\gamma'$, cube |
| 923 | $\gamma'$, octahedral | $\gamma'$, octahedral | $\gamma'$, octahedral | $\gamma$, octahedral | $\gamma'$, cube | $\gamma'$, cube |
| 973 | $\gamma'$, octahedral | $\gamma'$, octahedral | $\gamma'$, octahedral | $\gamma$, octahedral | $\gamma'$, cube | $\gamma'$, cube |
| 1023 | $\gamma$, octahedral | $\gamma'$, octahedral | $\gamma'$, octahedral | $\gamma$, octahedral | $\gamma'$, cube | $\gamma'$, cube |
| 1123 | $\gamma$, octahedral | $\gamma$, octahedral | $\gamma'$, cube | $\gamma'$, cube | $\gamma'$, cube | $\gamma'$, cube |

*5.4. Tension-compression asymmetry and slip system activity of CMSX-4*

The strength differential ($SD$) contours and the slip system activity contours in tension and compression are shown inside the standard stereographic triangle for CMSX-4 crystals for different temperatures in figure 10. As earlier, blue color indicates dominant octahedral slip at yield, while red color indicates dominant cube slip in the $\gamma'$ phase in the slip system activity plots. Additionally, the green color denotes octahedral slip in the $\gamma$ phase in these plots.

At 300 K, the value of $SD$ near the [0 0 1] and [$\bar{1}$ 1 1] poles is very small, while it is $\approx -0.3$ for regions near the [0 1 1] pole. The slip activity contours for the different regions reveal that while octahedral slip in the $\gamma'$ phase is responsible for the observed behavior near the [0 1 1] and [0 0 1] poles, cube slip in the $\gamma'$ phase is responsible for the yield behavior near the [$\bar{1}$ 1 1] pole. Note that slip (either octahedral or cube) in the $\gamma'$ phase is essentially representative of precipitate shear as the primary deformation mechanism at yield. These observations are also in agreement with experimental observations (Sengupta et al., 1994).

The $SD$ values (and the tension-compression asymmetry) increase at 600 K, although the trends in the slip system activity remain similar, with one difference: the region encompassing cube slip near the [$\bar{1}$ 1 1] pole has grown larger. Note that the regions in which cube slip is active show zero $SD$. This trend continues further at 800 K and 900 K, *albeit* with the appearance of octahedral slip in the $\gamma'$ phase near the [0 1 1] pole, thus leading to more negative values of $SD$. On the other hand, the $SD$ intensifies near the [0 0 1] pole up to 1000 K.

At higher temperatures, octahedral slip in the $\gamma$ phase dominates near the [0 0 1] pole. Since there are no non-Schmid stresses involved, the $SD$ is lower. At 1200 K, we observe minimal tension-compression asymmetry over the entire stereographic triangle. Although we do not have much experimental data to compare our $SD$ results with, the slip activity indicates that cube slip activation is prominent over a large region of the stereographic triangle. Dominance of cube slip for orientations near the [$\bar{1}$ 1 1] pole for all temperatures was also discussed by (Allan, 1995). Based on geometric considerations, the cube slip has the highest Schmid factor ($\approx 0.47$) for this loading orientation.

These observed slip system activities are consistent with those reported in table 6 for the different loading conditions. This may serve as a self-consistent validation of the model's ability to predict the underlying mechanisms, since the predicted yield stress values are in coherence with experiments (cf. section 5.3). There is limited experimental data available in the literature to support these model predictions. (Sengupta et al., 1994) note that precipitate shearing is operative up to about 1073 K for CMSX-4 crystals loaded along the [0 0 1] direction, beyond which Orowan looping (in the matrix) becomes active. Our model predicts a similar behavior for the [0 0 1] crystal over the temperature ranges of interest. In an experimental study, (Miner et al., 1986) found traces of cube slip in a Ni-based superalloy over a range of temperatures. Similar observations were also reported by (Westbrooke et al., 2005) in a single crystal Ni-based superalloy. Further, (Bettge and Österle, 1999; Österle et al., 2000) also found evidence of *macroscopic* cube slip traces by performing electron microscopy on [$\bar{1}$ 1 1] loaded single crystals. However, they also noted that the cube slip traces are presumably generated by coordinated zigzag dislocation motion on {1 1 1} planes. From a purely mathematical perspective, cube slip (with a Schmid factor $\approx 0.47$) is much more viable than octahedral slip (with a Schmid factor $\approx 0.27$) near the [$\bar{1}$ 1 1] pole.

Overall, the *SD* contours, combined with the slip system activity contours, allow us to understand the underlying mechanisms responsible for the observed yield behavior. The contours, which are similar to deformation mechanism maps, may prove valuable for assessing the design limits of these alloys. While previous studies have captured the anisotropic deformation of these materials, the underlying slip modes are not generally discussed in such detail. Specifically, these data may provide insights to the design engineers regarding crystal orientations that are preferable and those that should be avoided (orientations where deformation of the precipitate phase is dominant, for example) for loading the material; design limits of the material are often based on the yield stress.

We note that the inelastic deformation mechanisms may change with subsequent inelastic deformation. For example, (Daymond et al., 2007; Grant et al., 2011) have shown clear evidence of load transfer between the matrix and precipitate phases during the later stages of deformation in polycrystalline Ni-based superalloys. In these neutron diffraction studies, the $\gamma'$ phase was found to yield initially. With subsequent hardening of the $\gamma'$ phase, the $\gamma$ phase was found to become compliant to inelastic deformation during the later stages.

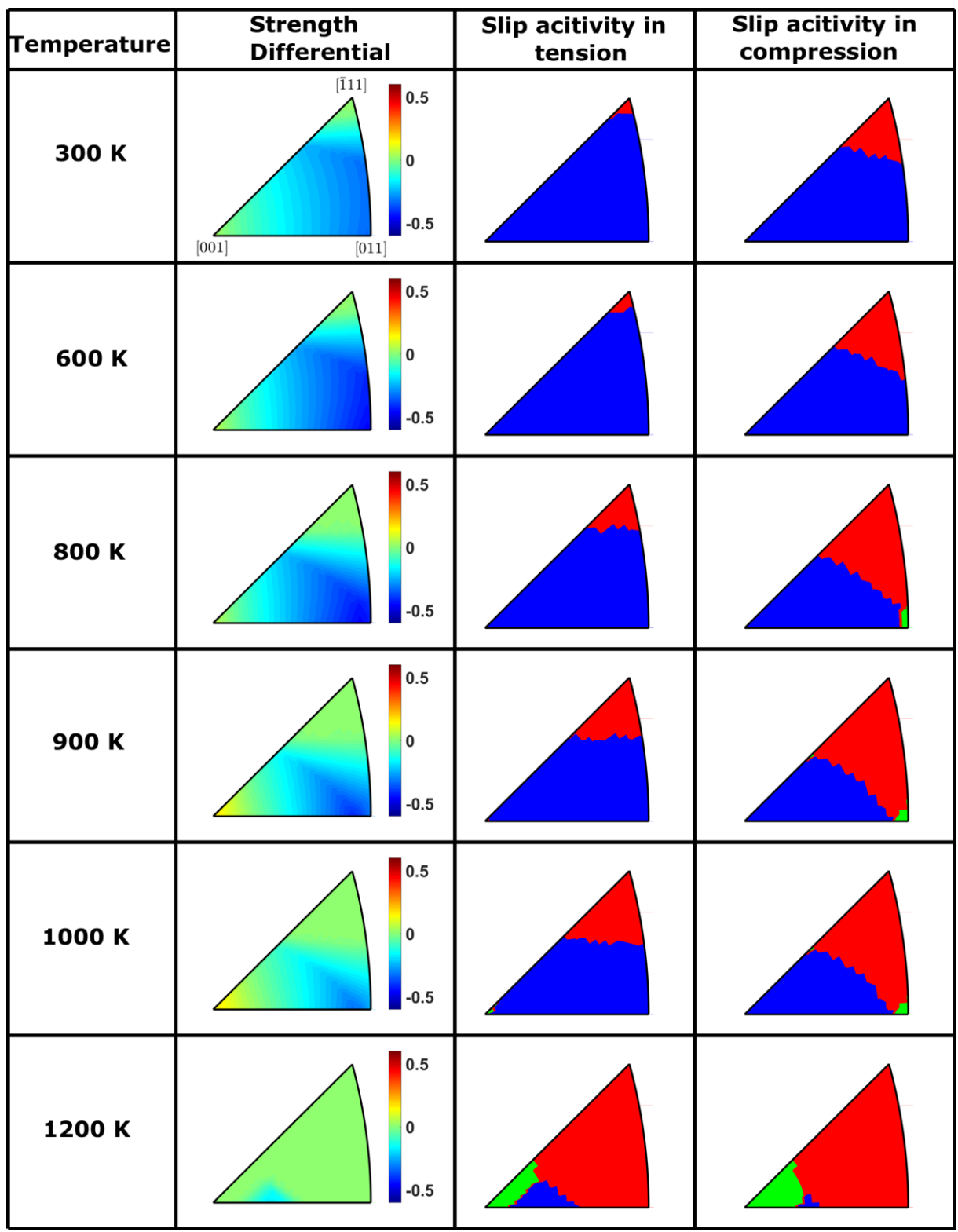

**Figure 10.** Prediction of strength differential (*SD*), and dominant slip activity in tension and compression for CMSX-4. In the slip activity plots, blue color indicates octahedral slip and red color indicates cube slip in the precipitate phase, while green color indicates octahedral slip in the matrix phase.

## 5.5. Parametric study of the exponential decay model for non-Schmid stresses

We have modeled the diminishing effect of the non-Schmid stresses by introducing an exponential decay function of the effective inelastic strain, $\bar{\varepsilon}_p^i$, i.e., $exp(-\bar{\varepsilon}_p^i/\varepsilon_0^i)$ (cf. eq. (16) in section 4.3). Here, $\varepsilon_0^i$ is a model parameter that may be calibrated to match the experiments. Phenomenologically, this parameter is representative of the effective inelastic strain at which the contribution of non-Schmid stresses to the driving stress decreases by about 37% of its initial contribution at yield, i.e., the point of commencement of inelastic deformation.

We have varied this model parameter, $\varepsilon_0^i$, to study its effect on the initial stress-strain response. Parametric study of the effect of this model parameter on the stress-strain response of $Ni_3Al$ and CMSX-4 crystals is shown in figure 11. As the value of $\varepsilon_0^i$ is increased from 0.002 to 0.01, we observe a monotonic decrease in the flow stress, beyond the point of initial yield for both materials. At higher applied deformation ($> 0.01$ strain), the stress-strain curves tend to converge, thus indicating that the contribution of the non-Schmid stresses to the driving force for dislocation glide decreases. Physically, this is representative of the constriction of the non-planar dislocation cores. While we have not attempted to calibrate the hardening response of the material to the experimental stress-strain data in the present work, we would like to point out that this drop in the flow stress subsequent to initial yield is indeed observed in Ni-based superalloys. The reader is referred to the experimental stress-strain curves reported in (Tinga et al., 2010; Vattré and Fedelich, 2011) for CMSX-4 and (Leidermark et al., 2009) for MD2 single crystal alloy. This non-Schmid decay model offers a phenomenological approach for representing such behavior.

While the results have only been shown for two specific loading and temperature conditions, we have performed this exercise for all loading orientations and temperatures for which the experimental data is available for $Ni_3Al$ and CMSX-4 (Allan, 1995; Heredia and Pope, 1991). Model predictions of the 0.2% offset stress are used to calculate the mean root mean square error over all orientations and loading types for both materials. These values are shown in table 7. The average normalised errors indicate that $\varepsilon_0^i = 0.005$ gives the best fit of the predicted 0.2% offset yield stress to the experimental yield stress data for $Ni_3Al$, and $\varepsilon_0^i = 0.01$ gives the best fit to the experimental yield stress data for CMSX-4. Note that the error associated with the prediction of the proportional limit stress does not change by changing this model parameter. This is due to the fact that the non-Schmid stresses starts decaying only after the point of initial yield. This model may be used for calibrating the predicted stress-strain response to the experimental data in future work.

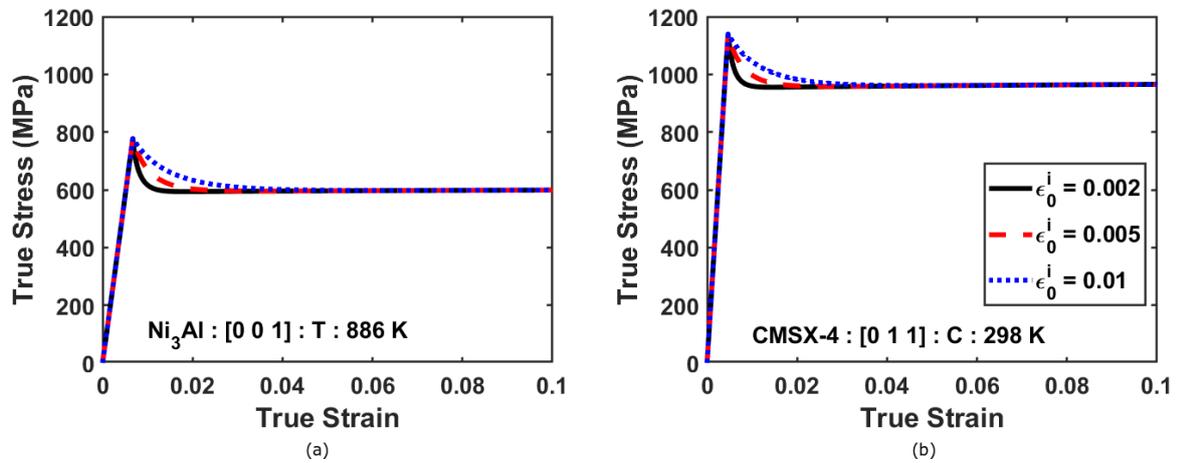

**Figure 11.** Effect of the decay parameter on the stress-strain response of (a) Ni$_3$Al, and (b) CMSX-4.

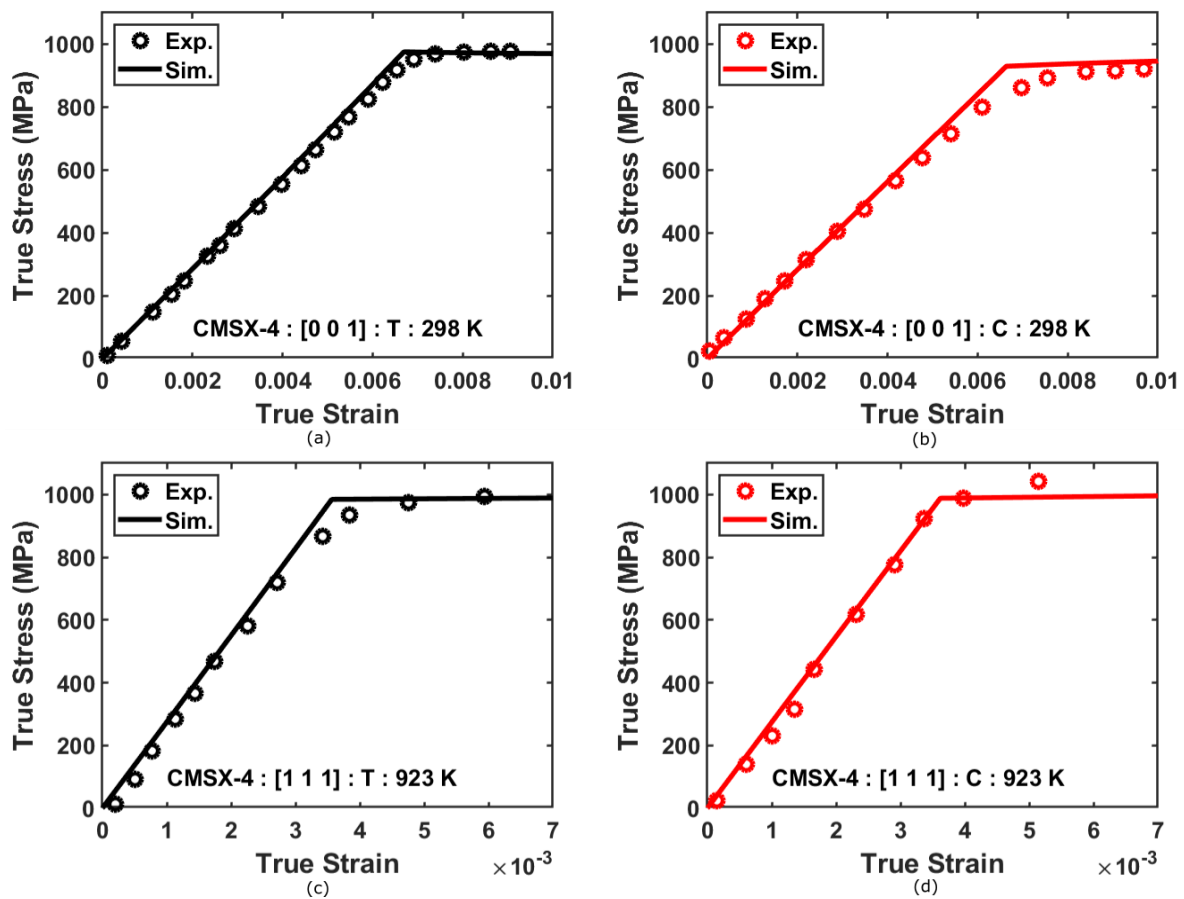

**Figure 12.** Comparison of simulated stress-strain response of CMSX-4 with the experimental data for (a) [0 0 1] loading in tension at 298 K, (b) [0 0 1] loading in compression at 298 K, (c) [$\bar{1}$ 1 1] loading in tension at 923 K, and (d) [$\bar{1}$ 1 1] loading in compression at 923 K. Note

that the experimental data (Allan, 1995) is available only up to small strain values.

We have presented model predictions up to 0.1 applied strain in Figure 11. However, due to the lack of experimental stress-strain data for higher applied strains in Allan (1995), we have not compared our model predictions with the same reference. There are other studies in the literature which give the experimental stress-strain response up to higher applied strains, for example see Vattre and Fedelich (2011) and Leidermark et al. (2010). We have not compared model predictions with these experiments because the measured yield stress itself is different from what has been reported in Allan (1995). Presumably, this is due to a difference in heat treatment histories (cf. Sengupta et al. (1994)). Nonetheless, it should be noted that little or no strain hardening was observed for CMSX-4 in the experimental studies of Vattre and Fedelich (2011) and Leidermark et al. (2010), as is also predicted by our model. This may also serve as a qualitative validation of our hardening parameters.

Finally, we present comparison of the simulated stress-strain response with the limited experimental data from (Allan, 1995) in Figure 12. We have compared model predictions for two different loading orientations, [0 0 1] and [$\bar{1}$ 1 1], at two different temperatures, 298 K and 923 K, for loading in tension and compression. The experimental data is available for strain values of 0.01 or less. As can be seen, there is a reasonable fit with the experimental data in all cases. Further, these results also serve as validation for predictions of the orientation- and temperature-dependent elastic modulus.

In future work, more rigorous validation of the model predictions of the post-yield deformation behavior may be performed, for example, by comparing texture evolution with the corresponding experimental data. The framework may also be extended, with small modifications, to explicitly account for individual phases in the geometric mesh and predict the phase-specific deformation. These phase-specific strains can then be compared with lattice strain evolution data available from experiments (cf. (Daymond et al., 2007; Grant et al., 2011)). In future work, biaxial simulations may also be performed to validate model predictions with the experimental data (cf. Allan (1995)).

**Table 7.** Effect of the decay parameter on the mean RMS error between CPFE predictions of the 0.2% offset yield stress and the experimental data over all orientations and temperatures for $Ni_3Al$ and CMSX-4.

| Material | Mean RMS error between CPFE predictions and experimental data over all orientations and temperatures | | |
|---|---|---|---|
| | $\varepsilon_0^i = 0.002$ | $\varepsilon_0^i = 0.005$ | $\varepsilon_0^i = 0.010$ |

|  | Proportional limit | 0.2% offset | 0.2% offset | 0.2% offset |
|---|---|---|---|---|
| Ni$_3$Al | 0.0992 | 0.125 | 0.1200 | 0.1215 |
| CMSX-4 | 0.0464 | 0.0607 | 0.0511 | 0.4790 |

*5.6. Connecting the constitutive models with the underlying physics*

This work has primarily relied on estimation of the CRSS for octahedral slip in Ni$_3$Al/$\gamma'$ phase directly from the experimental yield stress data (cf. section 3). While the physical processes governing inelastic deformation of the $\gamma'$ phase have been described in section 2, we have not attempted to connect the phenomenological model of the CRSS with these physical mechanisms. Note that the inelastic deformation of the $\gamma'$ phase in our constitutive model is simply representative of the precipitate shearing mechanism commonly discussed in the literature (Kear and Oblak, 1974; Kozar et al., 2009).

As mentioned earlier, the CRSS is representative of the strengthening caused by cross-slip dislocations and KW lock formation. The cross-slip process is in turn driven by the formation of APBs and hence the associated stacking fault energy. This energy, commonly referred as the Anti-Phase Boundary Energy (APBE), is material chemistry-, temperature-, strain rate- and slip system-dependent. One of the earliest calculations for the APBEs were made by (Flinn, 1960). In later works (Gorbatov et al., 2016; Veyssiere et al., 1985; Yoo et al., 1994), have also estimated the APBEs using lower scale atomistic calculations. A common finding in all these studies is that the APBE is higher for the {111} plane than the {001} plane. This promotes cross-slip to the {001} plane, thus reducing the energy of the system. Further, (Yu et al., 1994) noted that the APBE itself is temperature-dependent and that the driving force for KW lock formation increases with temperature, thus leading to an increase in the CRSS. A similar observation was also reported in (Manga et al., 2015).

Based on experiments performed on single crystal Ni-based superalloys, (Shah and Duhl, 1984) first proposed a relation between the CRSS and the APBE of the form:

$$CRSS = k \frac{\Delta \Gamma(T)}{b} + \text{other terms} \tag{27}$$

where $k$ is a material constant, $\Delta \Gamma(T)$ represents the temperature-dependent difference in APBE between the {111} and {001} planes and $b$ is the burgers vector magnitude. The other terms included a term describing the effect of the precipitate size, and a temperature dependent term denoting the effect of cross-slip mechanism, which restrict the shearing of precipitates. More physically based models were of course proposed by (Cuitiño and Ortiz, 1993; Paidar et al., 1984) (cf. eq. (1) and (2)). In another class of homogenized models (Ghorbanpour et al., 2017; Kozar et al., 2009; Shenoy et al., 2008), which do not consider the matrix and precipitate phases separately, the hardening due to precipitate is commonly given as a function of the square root of the $\Delta\Gamma(T)$.

Thus, the CRSS is a function of $\Delta\Gamma(T)$ for octahedral slip in the $\gamma'$ phase. With *a priori* knowledge (or via lower scale calculations) of the $\Delta\Gamma(T)$, the CRSS may be directly estimated for different alloys. These observations have further implications in alloy design, where the alloy composition itself may be modified to tune $\Delta\Gamma(T)$ and attain the desired mechanical properties (Collins and Stone, 2014).

Finally, the non-Schmid parameters are expected to depend primarily on the APB energy of the $\gamma'$ phase, which in turn is material chemistry dependent. Generally speaking, these should not depend on the volume fraction of the $\gamma'$ phase. However, the volume fraction of the $\gamma'$ phase may itself be dependent on the material chemistry and the processing history (Sajjadi et al., 2006; Wei et al., 2012). Hence, it may not be possible to entirely delineate the interdependence of these.

## 6. Conclusions

This work presents a Crystal Plasticity Finite Element (CPFE) framework for modeling the non-Schmid yield behavior of L1$_2$ type Ni$_3$Al crystals and two phase Ni-based superalloys. The framework can account for the effects of microstructure, specifically volume fraction, precipitate size and matrix channel width on the associated hardening mechanisms. Application of the framework to predict the yield stress is demonstrated over a large range of orientations and temperatures for Ni$_3$Al and CMSX-4. Significant contributions and findings of this work are summarized in the following:

1. A general algorithm based on the linear non-Schmid yield model (Qin and Bassani, 1992) is proposed to estimate the non-Schmid coefficients as a function of temperature. The non-Schmid model parameters for Ni$_3$Al have been directly estimated from the experimental data (Heredia and Pope, 1991). While previous studies (Dao and Asaro, 1993; Qin and Bassani, 1992) have proposed this type of a linear yield criterion for Ni₃Al, their application to the entire range of experimental yield stress data has not been shown. Here, we have demonstrated the same from 260 K to 1304 K and six crystal orientations covering different regions of the standard stereographic triangle.

2. The non-Schmid model parameters are directly utilized in the CPFE framework for predicting the orientation- and temperature-dependent yield of Ni₃Al crystals. Minimal calibration is required for the inelastic flow and hardening parameters in the crystal plasticity model once the non-Schmid parameters have been estimated, thus reducing the significant computational efforts generally involved in parameter estimation of CPFE models.

3. The non-Schmid model for Ni₃Al is extended to model deformation in the $\gamma'$ phase of a single crystal Ni-based superalloy, CMSX-4, with the implicit assumption that the deformation characteristics of Ni₃Al extend to the $\gamma'$ phase. Some modifications are needed to the non-Schmid parameters ($a_1$, $a_2$, and $a_3$) to fit to the experimental yield stress data for CMSX-4 (Allan, 1995) in order to account for the observed tension-compression asymmetry at lower temperatures. To the best of our knowledge, application of the linear non-Schmid model to simulate the yield behavior of CMSX-4 over this vast range of orientations and temperatures has not been demonstrated previously.

4. The calibrated linear non-Schmid model may be serve as a standalone analytical model for predicting the yield stress of Ni₃Al and CMSX-4 as a function of orientation and temperature.

5. The tension-compression asymmetry and slip activity are predicted for both Ni₃Al and CMSX-4 alloy over the standard stereographic triangle. These slip activity predictions provide valuable insights regarding the slip mechanisms governing yield, more specifically, the heterogeneity thereof with respect to crystal orientation and temperature. Such plots have not been reported elsewhere for Ni-based superalloys. The contours, which are similar to deformation mechanism maps, may prove valuable for assessing the design limits of these alloys.
    a) For Ni₃Al, we predict activation of octahedral slip at lower temperatures, while cube slip also gets activated in the regions near the $[\bar{1}\ 1\ 1]$ pole of the stereographic triangle at higher temperatures. Model predictions of the orientation- and temperature-dependent tension

compression asymmetry and the dominant slip mechanisms at yield are in concurrence with observation in the literature (Thornton et al., 1970).

b) Similar predictions are also made for CMSX-4. While yield is mostly dominated by slip in the precipitate phase at lower temperatures, slip in the matrix phase starts occurring at higher temperatures. The slip mechanisms predicted by our model are in concurrence with available experimental data (Daymond et al., 2007; Sengupta et al., 1994).

**Acknowledgements**

We would like to thank Prof. P.J. Guruprasad (IIT Bombay) for valuable feedback regarding the manuscript. Funding received from Siemens Technology and Services (Bangalore) is gratefully acknowledged. DR and AP also acknowledge funding received from the Aeronautics Research and Development Board (AR&DB) for this research via GTMAP project no. 1850.

**Appendix A: Slip systems considered for the crystal plasticity model**

**Table A1.** Octahedral and cube slip systems in Ni$_3$Al/$\gamma'$ phase (Qin and Bassani, 1992).

| $\alpha$ | Type | $(n^\alpha)[m^\alpha]$ | $(n^\alpha_{pe})[m^\alpha_{pe}]$ | $(n^\alpha_{se})[m^\alpha_{se}]$ | $(n^\alpha_{cb})[m^\alpha_{cb}]$ |
|---|---|---|---|---|---|
| 1 | Octahedral | $(111)[01\bar{1}]$ | $(111)[\bar{2}11]$ | $(1\bar{1}\bar{1})[\bar{2}1\bar{1}]$ | $(100)[01\bar{1}]$ |
| 2 | Octahedral | $(111)[\bar{1}01]$ | $(111)[1\bar{2}1]$ | $(\bar{1}1\bar{1})[\bar{1}2\bar{1}]$ | $(010)[\bar{1}01]$ |
| 3 | Octahedral | $(111)[1\bar{1}0]$ | $(111)[11\bar{2}]$ | $(\bar{1}\bar{1}1)[\bar{1}\bar{1}2]$ | $(001)[1\bar{1}0]$ |
| 4 | Octahedral | $(1\bar{1}\bar{1})[0\bar{1}1]$ | $(1\bar{1}\bar{1})[\bar{2}1\bar{1}]$ | $(111)[\bar{2}11]$ | $(100)[0\bar{1}1]$ |
| 5 | Octahedral | $(1\bar{1}\bar{1})[\bar{1}0\bar{1}]$ | $(1\bar{1}\bar{1})[12\bar{1}]$ | $(\bar{1}1\bar{1})[\bar{1}12]$ | $(0\bar{1}0)[\bar{1}0\bar{1}]$ |
| 6 | Octahedral | $(1\bar{1}\bar{1})[110]$ | $(1\bar{1}\bar{1})[1\bar{1}2]$ | $(\bar{1}1\bar{1})[\bar{1}12]$ | $(00\bar{1})[110]$ |
| 7 | Octahedral | $(\bar{1}1\bar{1})[011]$ | $(\bar{1}1\bar{1})[21\bar{1}]$ | $(\bar{1}\bar{1}1)[2\bar{1}1]$ | $(\bar{1}00)[011]$ |
| 8 | Octahedral | $(\bar{1}1\bar{1})[10\bar{1}]$ | $(\bar{1}1\bar{1})[\bar{1}2\bar{1}]$ | $(111)[1\bar{2}1]$ | $(010)[10\bar{1}]$ |
| 9 | Octahedral | $(\bar{1}1\bar{1})[\bar{1}\bar{1}0]$ | $(\bar{1}1\bar{1})[\bar{1}12]$ | $(1\bar{1}\bar{1})[1\bar{1}2]$ | $(00\bar{1})[\bar{1}\bar{1}0]$ |
| 10 | Octahedral | $(\bar{1}\bar{1}1)[0\bar{1}\bar{1}]$ | $(\bar{1}\bar{1}1)[2\bar{1}1]$ | $(\bar{1}1\bar{1})[21\bar{1}]$ | $(\bar{1}00)[0\bar{1}\bar{1}]$ |
| 11 | Octahedral | $(\bar{1}\bar{1}1)[101]$ | $(\bar{1}\bar{1}1)[\bar{1}21]$ | $(1\bar{1}\bar{1})[1\bar{1}2]$ | $(0\bar{1}0)[101]$ |
| 12 | Octahedral | $(\bar{1}\bar{1}1)[\bar{1}10]$ | $(\bar{1}\bar{1}1)[\bar{1}\bar{1}\bar{2}]$ | $(111)[11\bar{2}]$ | $(001)[\bar{1}10]$ |
| 13 | Octahedral | $(111)[0\bar{1}1]$ | $(111)[\bar{2}11]$ | $(1\bar{1}\bar{1})[\bar{2}1\bar{1}]$ | $(100)[0\bar{1}1]$ |

| 14 | Octahedral | $(111)[10\bar{1}]$ | $(111)[1\bar{2}1]$ | $(\bar{1}1\bar{1})[\bar{1}\bar{2}\bar{1}]$ | $(010)[10\bar{1}]$ |
|---|---|---|---|---|---|
| 15 | Octahedral | $(111)[\bar{1}10]$ | $(111)[11\bar{2}]$ | $(\bar{1}\bar{1}1)[\bar{1}\bar{1}\bar{2}]$ | $(001)[\bar{1}10]$ |
| 16 | Octahedral | $(1\bar{1}\bar{1})[01\bar{1}]$ | $(1\bar{1}\bar{1})[\bar{2}\bar{1}\bar{1}]$ | $(111)[\bar{2}11]$ | $(100)[01\bar{1}]$ |
| 17 | Octahedral | $(1\bar{1}\bar{1})[101]$ | $(1\bar{1}\bar{1})[12\bar{1}]$ | $(\bar{1}1\bar{1})[\bar{1}12]$ | $(0\bar{1}0)[101]$ |
| 18 | Octahedral | $(1\bar{1}\bar{1})[\bar{1}\bar{1}0]$ | $(1\bar{1}\bar{1})[1\bar{1}2]$ | $(\bar{1}1\bar{1})[\bar{1}12]$ | $(00\bar{1})[\bar{1}\bar{1}0]$ |
| 19 | Octahedral | $(\bar{1}1\bar{1})[0\bar{1}\bar{1}]$ | $(\bar{1}1\bar{1})[21\bar{1}]$ | $(\bar{1}\bar{1}1)[2\bar{1}1]$ | $(\bar{1}00)[0\bar{1}\bar{1}]$ |
| 20 | Octahedral | $(\bar{1}1\bar{1})[\bar{1}01]$ | $(\bar{1}1\bar{1})[\bar{1}\bar{2}\bar{1}]$ | $(111)[1\bar{2}1]$ | $(010)[\bar{1}01]$ |
| 21 | Octahedral | $(\bar{1}1\bar{1})[110]$ | $(\bar{1}1\bar{1})[\bar{1}12]$ | $(1\bar{1}\bar{1})[1\bar{1}2]$ | $(00\bar{1})[110]$ |
| 22 | Octahedral | $(\bar{1}\bar{1}1)[011]$ | $(\bar{1}\bar{1}1)[2\bar{1}1]$ | $(\bar{1}1\bar{1})[21\bar{1}]$ | $(\bar{1}00)[011]$ |
| 23 | Octahedral | $(\bar{1}\bar{1}1)[\bar{1}0\bar{1}]$ | $(\bar{1}\bar{1}1)[\bar{1}21]$ | $(1\bar{1}\bar{1})[1\bar{1}2]$ | $(0\bar{1}0)[\bar{1}0\bar{1}]$ |
| 24 | Octahedral | $(\bar{1}\bar{1}1)[1\bar{1}0]$ | $(\bar{1}\bar{1}1)[\bar{1}\bar{1}\bar{2}]$ | $(111)[11\bar{2}]$ | $(001)[1\bar{1}0]$ |
| 25 | Cube | $(100)[011]$ | - | - | - |
| 26 | Cube | $(100)[01\bar{1}]$ | - | - | - |
| 27 | Cube | $(100)[01\bar{1}]$ | - | - | - |
| 28 | Cube | $(010)[01\bar{1}]$ | - | - | - |
| 29 | Cube | $(001)[110]$ | - | - | - |
| 30 | Cube | $(001)[1\bar{1}0]$ | - | - | - |

**Table A2.** Octahedral slip systems in the $\gamma$ phase.

| $\alpha$ | Type | $(n^{\alpha})[m^{\alpha}]$ |
|---|---|---|
| 1 | Octahedral | $(111)[01\bar{1}]$ |
| 2 | Octahedral | $(111)[\bar{1}01]$ |
| 3 | Octahedral | $(111)[1\bar{1}0]$ |
| 4 | Octahedral | $(1\bar{1}\bar{1})[0\bar{1}1]$ |
| 5 | Octahedral | $(1\bar{1}\bar{1})[\bar{1}0\bar{1}]$ |
| 6 | Octahedral | $(1\bar{1}\bar{1})[110]$ |
| 7 | Octahedral | $(\bar{1}1\bar{1})[011]$ |

| 8  | Octahedral | $(\bar{1}1\bar{1})[10\bar{1}]$ |
| 9  | Octahedral | $(\bar{1}1\bar{1})[\bar{1}\bar{1}0]$ |
| 10 | Octahedral | $(\bar{1}\bar{1}1)[0\bar{1}\bar{1}]$ |
| 11 | Octahedral | $(\bar{1}\bar{1}1)[101]$ |
| 12 | Octahedral | $(\bar{1}\bar{1}1)[\bar{1}10]$ |

**Appendix B: Parametric study of the non-Schmid coefficients on the tension-compression asymmetry**

Figure A1 shows the effect of variation of the non-Schmid coefficients on the strength differential ($SD$) (eq. (26)) over the entire stereographic triangle for octahedral slip in Ni$_3$Al. For this study, the non-Schmid coefficients $a_1$, $a_2$ and $a_3$ are individually assigned values of either $-0.5$ or $0.5$, while assigning a zero value to the other non-Schmid coefficients. It is easier to interpret these results in terms of the Schmid and non-Schmid factors at the corners of the stereographic triangle for the different loading types. These are listed for uniaxial loading in table A3. Note that the non-Schmid factors corresponding to $\tau_{pe}$ and $\tau_{se}$ do not change in sign with the loading type (tension/compression), while the Schmid factor and the non-Schmid factor corresponding to $\tau_{cb}$ changes sign.

The individual effects of $a_1$ and $a_2$ appear to be qualitatively similar. For a negative value of either $a_1$ or $a_2$, we observe a region of positive $SD$ near [0 0 1] pole while a region of negative $SD$ values near [0 1 1] and [$\bar{1}$ 1 1] poles. On changing the value of the parameter to 0.5, the observed trend also reverses. These coefficients influence the $\tau_{pe}$ and $\tau_{se}$, which are responsible for the constriction of the Shockley partials in one sense of loading, while extending the partials in the other sense. A negative value of either $a_1$ or $a_2$ represents the former, while a positive value represents the latter case.

There is no tension-compression asymmetry observed for both $a_3 = -0.5$ and $a_3 = 0.5$. Physically, this parameter is associated with $\tau_{cb}$ acting on the cube plane, which facilities cross-slip. The sign of the non-Schmid factor associated with this coefficient is always the same as the Schmid factor, hence the zero $SD$.

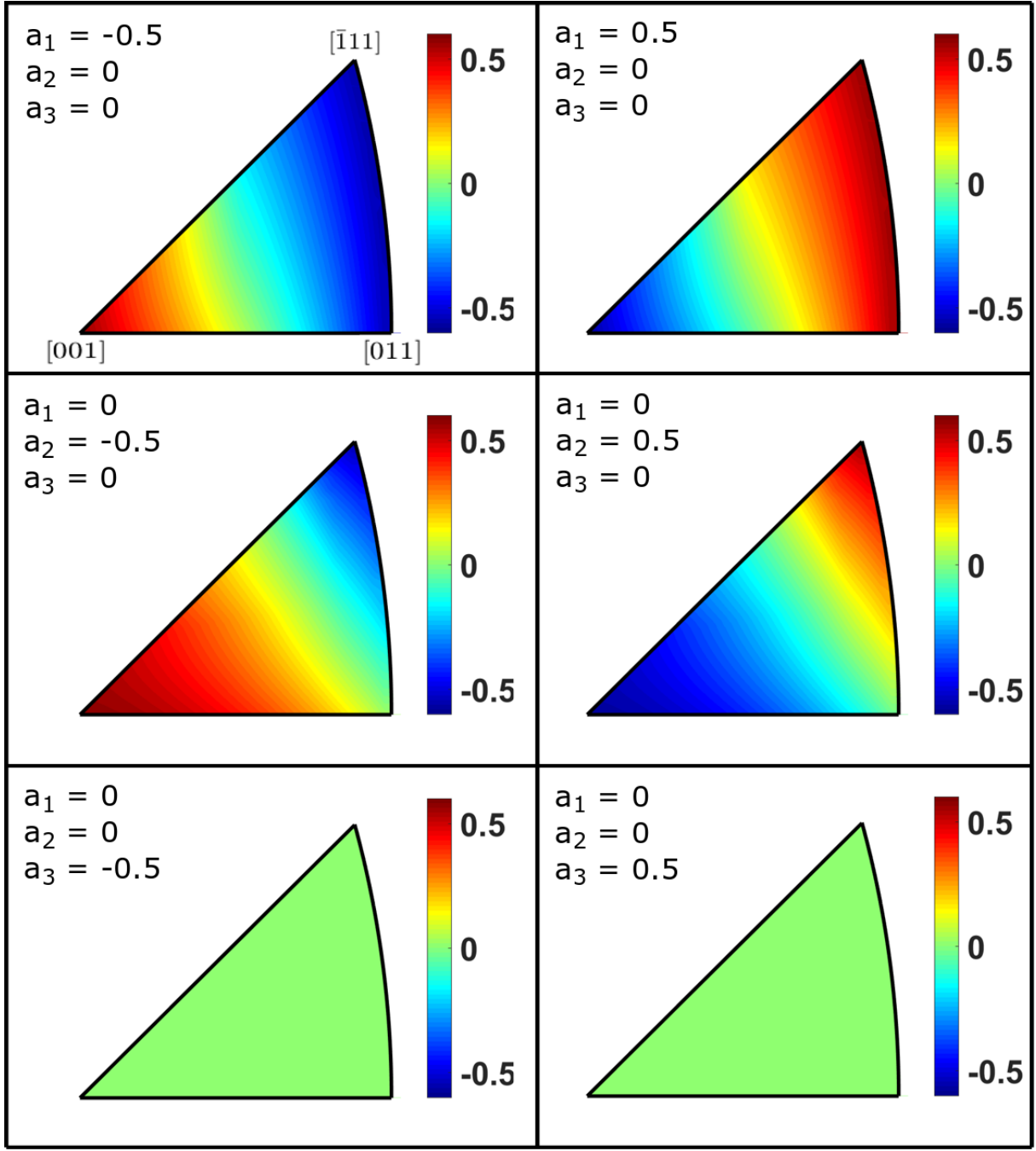

**Figure A1.** Parametric study of the effect of the non-Schmid parameters on the tension-compression asymmetry of Ni$_3$Al crystals assuming only octahedral slip.

**Table A3.** Schmid and non-Schmid factors for different loading orientations and types for uniaxial loading along the 33 direction.

| Loading orientation/type | $m^\alpha \cdot \sigma \cdot n^\alpha$ | $m^\alpha_{pe} \cdot \sigma \cdot n^\alpha_{pe}$ | $m^\alpha_{se} \cdot \sigma \cdot n^\alpha_{se}$ | $m^\alpha_{cb} \cdot \sigma \cdot n^\alpha_{cb}$ |
|---|---|---|---|---|
| [0 0 1]/T | 0.4082 | 0.2357 | 0.2357 | 0 |
| [0 0 1]/C | -0.4082 | 0.2357 | 0.2357 | 0 |
| [0 1 1]/T | 0.4082 | -0.2357 | 0 | 0.3535 |
| $[0\ 1\ 1]/C$ | -0.4082 | -0.2357 | 0 | -0.3535 |
| $[\bar{1}\ 1\ 1]/T$ | 0.2722 | -0.1571 | -0.1571 | 0.4714 |
| $[\bar{1}\ 1\ 1]/C$ | -0.2722 | -0.1571 | -0.1571 | -0.4714 |

**Appendix C: Solution of the non-Schmid parameters, $a_1, a_2, a_3$ and $\tau^\alpha_{0,oct}$ for Ni$_3$Al single crystal by using least square regression method**

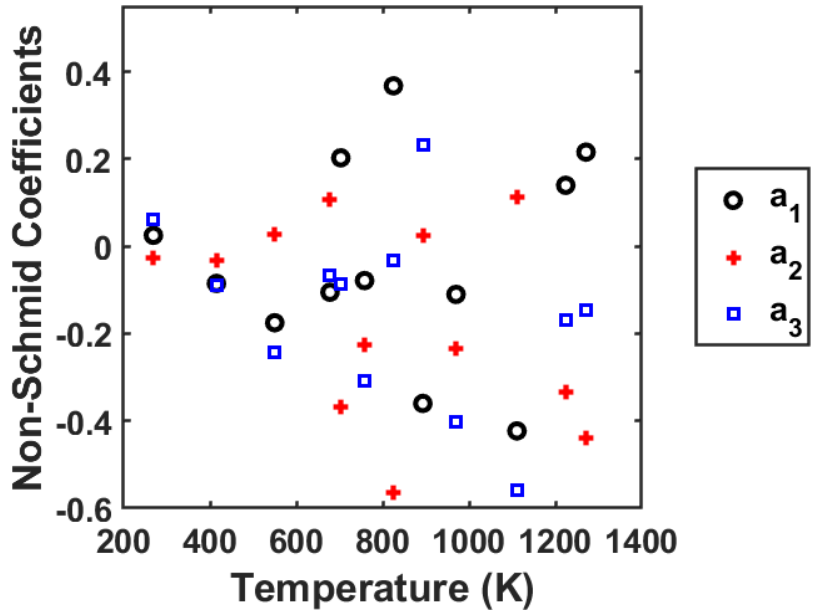

**Figure A2.** Variation of non-Schmid coefficients, $a_1$, $a_2$ and $a_3$ as a function temperature for octahedral slip in Ni$_3$Al estimated using least square regression method.

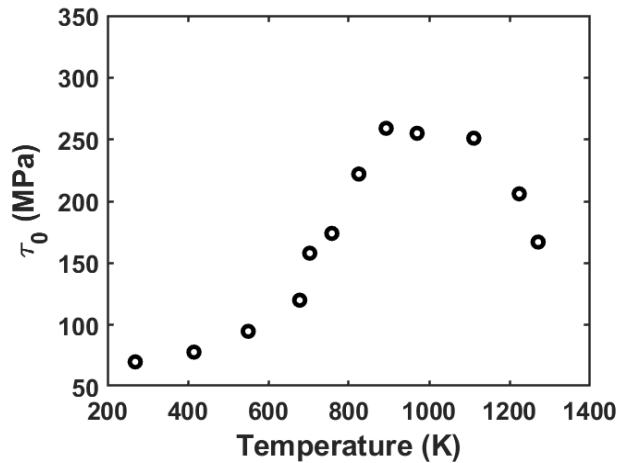

**Figure A3.** Variation of CRSS as a function of temperature for octahedral slip in Ni$_3$Al estimated using least square regression method.

**Appendix D: Model predictions of the proportional limit yield stress and the 0.2% offset yield stress for Ni$_3$Al**

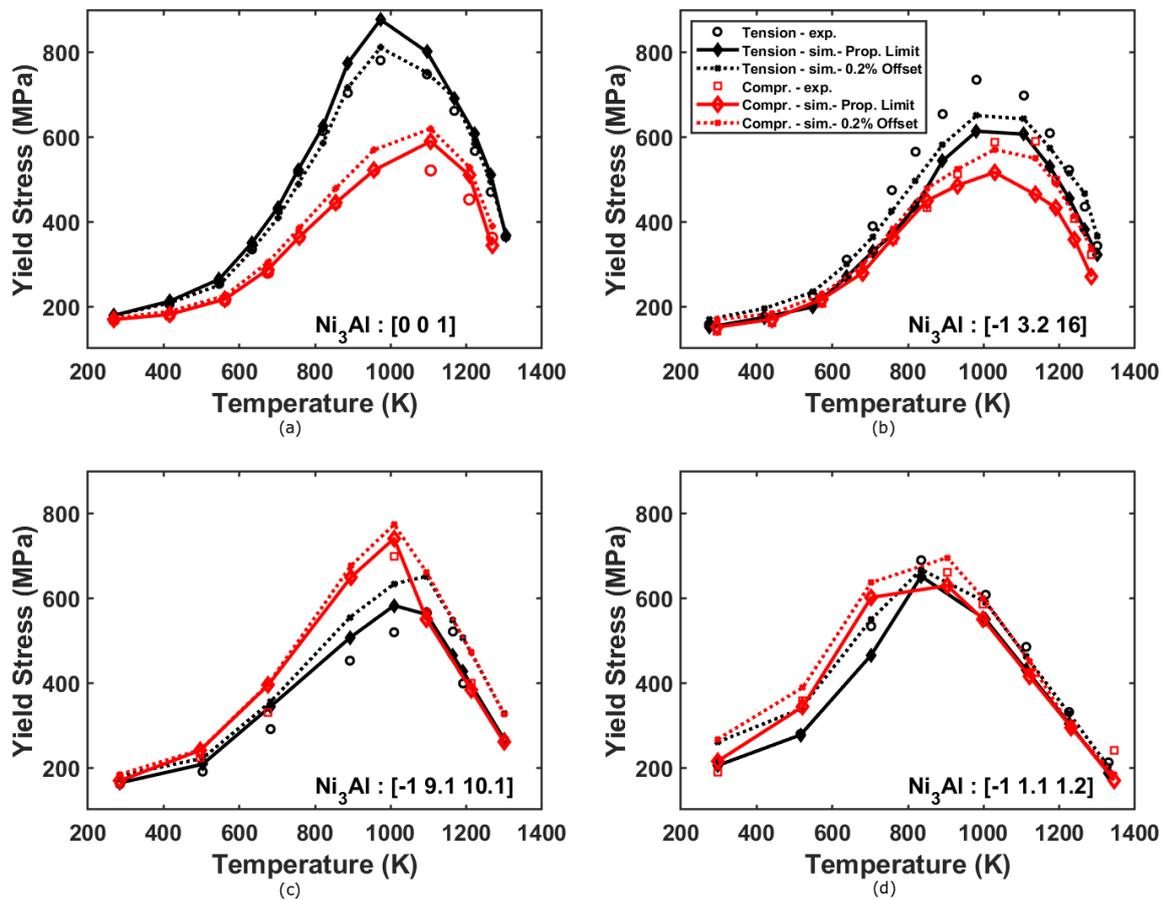

**Figure A4.** Model predictions of the proportional limit yield stress and the 0.2% offset yield stress for Ni$_3$Al single crystals compared with experiments.